\documentclass[aps,twocolumn,nofootinbib,showpacs,prd,aps,superscriptaddress,10pt]{revtex4-2}
\usepackage[dvips]{graphicx}
\usepackage[english]{babel}
\usepackage[T1]{fontenc}
\usepackage{mathrsfs}
\usepackage[tbtags]{amsmath}
\usepackage{amssymb}
\usepackage{amsxtra}
\usepackage{xcolor}
\usepackage{amsopn}
\usepackage{latexsym}
\usepackage[mathcal]{eucal}
\usepackage{mathtools}

\newcommand{\BE}{\begin{equation}}
\newcommand{\EE}{\end{equation}}
\newcommand{\BA}{\begin{align}}
\newcommand{\EA}{\end{align}}
\newcommand{\Tr}{\mathrm Tr}
\newcommand{\nn}{\nonumber}

\newcommand{\kkk}{ \frac{{\rm d}^4k}{(2\pi)^4}}

\newcommand{\qqqE}{ \frac{{\rm d}_{E}^4q}{(2\pi)^4}}

\newcommand{\Rerm}{\mathop{\rm Re}}

\renewcommand{\Im}{\mathop{\rm Im}}

\begin{document}
%\begin{document}

\newcommand{\unictandinfnct}{Dipartimento di Fisica e Astronomia dell'Universit\`a di Catania, INFN Sezione di Catania, Via S.Sofia 64, I-95123 Catania, Italy}
\newcommand{\kulak}{KU Leuven Campus Kulak Kortrijk, Department of Physics, Etienne Sabbelaan 53 bus 7657, 8500 Kortrijk, Belgium}

\title{The Nielsen identities in screened theories}

\author{Fabio Siringo}\email{fabio.siringo@ct.infn.it}\affiliation{\unictandinfnct}
\author{Giorgio Comitini}\email{giorgio.comitini@dfa.unict.it}\affiliation{\unictandinfnct}\affiliation{\kulak}

\date{\today}

\begin{abstract}
One-loop explicit expressions are derived for the gluon Nielsen identity in the formalism of the screened massive expansion
for Yang-Mills theory. The gauge-parameter-independence of the poles and residues is discussed in a
strict perturbative context and, more generally, in extended resummation schemes. No exact formal proof was reached by the approximate resummation schemes, but some evidence is gathered in favor of an exact invariance of the phase, consistently with previous numerical studies.
\end{abstract}

%\pacs{12.38.Aw, 12.38.Bx, 14.70.Dj, 12.38.Lg}

%12.38.Bx       Perturbative calculations
%12.38.Lg	Other nonperturbative calculations (QCD)
%12.38.Aw	General properties of QCD (dynamics, confinement, etc.)
%14.70.Dj	Gluons
%11.15.Tk       Other nonperturbative techniques (gauge field theories)

%12.38.Gc	Lattice QCD calculations (see also 11.15.Ha Lattice gauge theory)
%11.10.Ef       Field Theory: Lagrangian and Hamiltonian approaches
%11.15.-q	Gauge field theories
%12.20.-m       QED
%11.15.Bt       General properties of perturbation theory (gauge theory)
%12.38.-t	Quantum chromodynamics

\maketitle
\newpage

\section{Introduction}
Confinement and dynamical mass generation are among the most important open problems of
contemporary physics. The quantum field theories which describe the interactions of quarks and gluons,
QCD and pure Yang-Mills theory, are believed to be fully consistent theories, at all scales, 
containing a dynamical cutoff in the IR. But unfortunately, we are still far from a full understanding of the confinement
mechanism which seems to be somehow related to the dynamical generation of almost all the mass 
which is observed in the universe. 
Lattice and continuous studies \cite{cornwall,bernard,dono,philip,olive,aguilar04,papa15b,
cucch07,cucch08,cucch08b,cucch09,bogolubsky,
olive09,dudal,binosi12,olive12,burgio15,duarte,aguilar8,aguilar10,aguilar14,papa15,fischer2009,
huber14,huber15g,huber15b,pawlowski10,pawlowski10b,pawlowski13,
varqcd,genself,watson10,watson12,rojas,var,qed,higher,
reinhardt04,reinhardt05,reinhardt14,
journey,varT,cucch09b,cucch09c,cucch10,bicudo15,roberts17}
have ruled out the existence of a Landau pole and supported the
existence of a finite coupling, which is not too large even  deep in the IR. On the other hand, 
the important role of the analytic properties of the Green functions, and their relation with the dynamics, is
still largely unexplored because of the breakdown of ordinary perturbation theory and of the lack of
alternative analytical tools in the continuous.

Quite recently, by a change of the expansion point, a new perturbative approach 
has been developed~\cite{ptqcd,ptqcd2,analyt,scaling,xigauge,RG,beta,ghost,damp,thermal,quark},
a {\it screened massive expansion} which is perfectly sound in the IR and has the usual merits of ordinary perturbation theory:
calculability, analytical outputs and a manifest description of the analytic properties in the complex plane.
The method gives direct and quantitative predictions for the poles of the gluon propagator which appear as complex conjugated
polar singularities~\cite{analyt,damp,thermal}.

The existence of complex conjugated poles was predicted by several models, 
like the refined Gribov-Zwanziger model~\cite{dudal08,capri,dudal10,dudal11,dudal18}, and their
deep effects on the dynamical properties of the gluon and on the confinement of color have been discussed by many authors~\cite{stingl,stingl96,kondo,kondo2}.
Moreover, a pair of complex conjugated poles invalidates the existence 
of the K\"allen-Lehmann representation~\cite{dispersion}, raises
important questions on the correct analytic continuation of the gluon propagator and jeopardizes the analytic properties of 
Dyson-Schwinger equations, unless some compensation arises from the unknown structure of the exact vertices~\cite{pawlowski22}.

On the physical meaning of the complex poles there are different, contrasting, opinions.
A recent formal approach~\cite{kondo2} has embraced the view that the complex poles would emerge from unphysical zero-norm states which
should be removed from the Hilbert space, giving rise to a confinement mechanism. However, the formal removal of quarks and gluons from the
physical states does not seem a satisfying solution for the problem of confinement, which would miss a more physical and dynamical
explanation. Moreover, according to that formal approach, 
the analytic continuation of the gluon propagator does not exist~\cite{kondo2}, raising
serious issues on the physical content of the theory.

A more physical approach~\cite{stingl96,damp} relies on the idea that quarks and gluons do exist, as internal degrees of freedom of the theory, but their phenomenological appearance is damped by a very short lifetime which confines them. In that view, the complex poles would play
a physical dynamical role in confinement, besides having to do with the dynamical mass which is observed in the IR.
That approach is corroborated by the discovery that, not only the poles, but even the phases of the complex residues 
appear to be gauge-parameter-independent.
The whole principal part of the gluon propagator seems to be gauge invariant: the phase of the residue is found to change less than $3\cdot 10^{-3}$ when
the gauge parameter goes from $\xi=0$ to $\xi=1.2$~\cite{xigauge}.
On the other hand, in a modified  K\"allen-Lehmann representation, the phase of the residues could be the direct consequence of a complex spectrum, and the invariance of the phase could be itself related to the gauge invariance of the spectrum.  
The same principal part seems to give the main contribution to a dimension-two condensate~\cite{boucaud} 
and to the short-range linear raising potential
which emerges from the Fourier transform of the propagator at  the leading order.
Thus, many arguments  would favor the gauge invariance of the phase of the residues if the gluons are believed to be
confined but still physical degrees of freedom.

From a formal point of view, a proof of gauge invariance would require 
the study of the Nielsen identities~\cite{nielsen,kobes90,breck}, exact
identities which determine the gauge dependence of the propagator in a covariant gauge. The identities are a direct consequence of the Becchi-Rouet-Stora-Tyutin (BRST)  symmetry which is displayed by the Faddeev-Popov Lagrangian of QCD and Yang-Mills theories. There is a growing interest in the role of the Nielsen identities for determining the properties of 
the propagators in a generic covariant gauge~\cite{papaxi} and for their explicit numerical evaluation~\cite{huber21}.

In this paper, the Nielsen identity for the gluon propagator is evaluated  by an explicit one-loop calculation in the framework of a 
screened perturbative expansion. Here, our primary interest is in the screened massive expansion, 
but the explicit one-loop expressions might be
useful for other screened theories, 
like the Curci-Ferrari model~\cite{tissier10,tissier11,tissier14,serreau,reinosa,pelaez,pelaez21}.
Moreover, the result can be pushed beyond a strict one-loop expansion by some resummation of infinite classes of graphs.

Because of the soft breaking of BRST which occurs in the screened expansion at any fixed order, the Nielsen identities are not expected to be fulfilled at one loop in our framework. Nonetheless, it is instructive to explore how the results change when going from the strictly perturbative expressions to those obtained by an approximate resummation of the internal gluon lines. The detailed study of the analytic properties of the latter seems to suggest that the phase might be exactly invariant, as expected both numerically~\cite{xigauge} and by physical arguments -- if the gluon principal part is to play a genuine physical role on the dynamics of the strong interactions. Thus, enforcing the pole and phase invariance turns out to be a consistent criterion 
for the optimization of the screened expansion from first principles, as was done in~\cite{xigauge} with remarkably good results.

Besides the perturbative, partially resummed, context, we are still not able to provide an exact formal proof for the invariance of the phases of the residues.
 
This paper is organized as follows: the massive screened expansion is briefly reviewed in Sec. II, 
in order to fix the notation; in Sec. III the Nielsen identity for the gluon propagator is derived and its relation with the polarization function is discussed; in Sec. IV the explicit one-loop expression of the identity is derived by the screened expansion; in Sec. V the one-loop result is discussed both in the perturbative context and by using different resummation schemes.
A detailed account of the explicit steps leading to the evaluation of the one-loop graphs is reported in the Appendix.

\section{The Screened expansion}

The massive, {\it screened} expansion was first developed in Refs.~\cite{ptqcd,ptqcd2} and related to the Gaussian effective potential
in Refs.~\cite{journey,varT}. It is based on a change of the expansion
point of ordinary perturbation theory.

In the pure gauge sector, the gauge-fixed Lagrangian can be written as
\BE
{\cal L}={\cal L}_{YM}+{\cal L}_{fix}+{\cal L}_{FP}\,,
\label{Ltot}
\EE
where ${\cal L}_{YM}$ is the Yang-Mills term
\BE
{\cal L}_{YM}=-\frac{1}{2} \Tr\left(  \hat F_{\mu\nu}\hat F^{\mu\nu}\right),
\EE
${\cal L}_{FP}$ is the ghost term
arising from the Faddeev-Popov determinant and
${\cal L}_{fix}$ is a covariant gauge-fixing term,
\BE
{\cal L}_{fix}=-\frac{1}{\xi} \Tr\left[(\partial_\mu \hat A^\mu)(\partial_\nu \hat A^\nu)\right]\,.
\label{fix}
\EE

Usually, the total action is split as $S_{tot}=S_0+S_I$ where the quadratic part
can be written as
\begin{align}
S_0&=\frac{1}{2}\int A_{a\mu}(x)\delta_{ab} {\Delta_0^{-1}}^{\mu\nu}(x,y) A_{b\nu}(y) {\rm d}^4 x\,{\rm d}^4 y \nn \\
&+\int \bar c_a(x) \delta_{ab}{{\cal G}_0^{-1}}(x,y) c_b (y) {\rm d}^4 x\, {\rm d}^4 y\,,
\label{S0}
\end{align}
while the interaction contains the three terms
\BE
S_I=\int{\rm d}^dx \left[ {\cal L}_{gh} + {\cal L}_3 +   {\cal L}_4\right]\,,
\label{SI}
\EE
which read
\begin{align}
{\cal L}_3&=-g  f_{abc} (\partial_\mu A_{a\nu}) A_b^\mu A_c^\nu\,,\nn\\
{\cal L}_4&=-\frac{1}{4}g^2 f_{abc} f_{ade} A_{b\mu} A_{c\nu} A_d^\mu A_e^\nu\,,\nn\\
{\cal L}_{gh}&=-g f_{abc} (\partial_\mu \bar c_a) c_b A_c^\mu\,.
\label{Lint}
\end{align}
In Eq.~(\ref{S0}), $\Delta_0$ and ${\cal G}_0$ are the standard free-particle propagators for
gluons and ghosts, respectively,  and their Fourier transforms are
\begin{align}
{\Delta_0}^{\mu\nu} (p)&=\Delta_0(p)\left[t^{\mu\nu}(p)
+\xi \ell^{\mu\nu}(p) \right]\,,\nn\\
\Delta_0(p)&=\frac{1}{-p^2}\,, \qquad {{\cal G}_0} (p)=\frac{1}{p^2}\,,
\label{D0}
\end{align}
having used the transverse and longitudinal projectors 
\BE
t_{\mu\nu} (p)=g_{\mu\nu}  - \frac{p_\mu p_\nu}{p^2}\,,\quad
\ell_{\mu\nu} (p)=\frac{p_\mu p_\nu}{p^2}\,.
\label{tl}
\EE

The screened massive expansion is obtained by a change of the quadratic expansion point, adding a transverse mass term to the quadratic
part of the action and subtracting it again from the interaction, thus leaving the total action
unchanged\footnote{This is actually done after renormalizing the Lagrangian. The details can be found e.g. in \cite{RG}.}. We add and subtract the action term 
\BE
\delta S= \frac{1}{2}\int A_{a\mu}(x)\>\delta_{ab}\> \delta\Gamma^{\mu\nu}(x,y)\>
A_{b\nu}(y) {\rm d}^4\, x{\rm d}^4y\,,
\label{dS1}
\EE
where the vertex function $\delta\Gamma$ is a shift of the inverse propagator,
\BE
\delta \Gamma^{\mu\nu}(x,y)=
\left[{\Delta_m^{-1}}^{\mu\nu}(x,y)- {\Delta_0^{-1}}^{\mu\nu}(x,y)\right]\,,
\label{dG}
\EE
and ${\Delta_m}^{\mu\nu}$ is a new massive free-particle propagator,
\begin{align}
{\Delta_m}^{\mu\nu} (p)&=
\Delta_m(p)\,t^{\mu\nu}(p)  
+\frac{\xi}{-p^2}\ell^{\mu\nu}(p)\,,
\label{Deltam}
\end{align}
with a massive transverse component
\BE
\Delta_m(p)=\frac{1}{-p^2+m^2}\,.
\label{DeltamT}
\EE
Adding that action term is equivalent to substituting the new massive propagator ${\Delta_m}^{\mu\nu}$ for the 
old massless one ${\Delta_0}^{\mu\nu}$ in the quadratic part.

Of course, in order to leave the total action unaffected by the change, we must include  the new interaction vertex, $\delta\Gamma$, among the standard interaction terms.
Dropping all color indices in the diagonal matrices and
inserting Eqs.~(\ref{D0}) and (\ref{Deltam}) into Eq.~(\ref{dG}), the vertex is just the transverse mass shift
of the quadratic part,
\BE
\delta \Gamma^{\mu\nu} (p)=m^2 t^{\mu\nu}(p)\,. 
\label{dG2}
\EE

The proper gluon polarization $\Pi$ and ghost self energy $\Sigma$ can then be evaluated, order by order, by perturbation theory.
In all Feynman graphs the internal gluon lines are replaced by the massive free-particle propagator ${\Delta_m}^{\mu\nu}$ while the new two-point vertex can be regarded as a new (transverse) 
mass counterterm, $\delta \Gamma^{\mu\nu}$, to be inserted in order to compensate the shift of the quadratic term in the action. The new two-point vertex is usually represented by a cross, like other counterterms, and we will refer to the 
graphs with one ore more crosses as {\it crossed} graphs.

Since the total gauge-fixed FP Lagrangian is not modified and because of gauge invariance,
the longitudinal polarization is known exactly and is zero, so that the total polarization
is transverse,
\BE
\Pi^{\mu\nu}(p)=\Pi(p)\, t^{\mu\nu}(p)\,,
\label{pol}
\EE
and the (exact) dressed propagators read
\begin{align}
\Delta_{\mu\nu}(p)&=\Delta (p)\,t_{\mu\nu}(p)+\Delta^L (p)\,\ell^{\mu\nu}(p)\,,\nn\\
{\cal G}^{-1}(p)&=p^2-\Sigma (p)\,,
\end{align}
where the transverse and longitudinal parts are given by
\begin{align}
{\Delta}^{-1} (p)&=-p^2+m^2-\Pi(p)\,,\nn\\
{\Delta^L} (p)&=\frac{\xi}{-p^2}\,.
\label{DTL}
\end{align}

At tree level, the polarization is given by the counterterm $\delta \Gamma$ of Eq.~(\ref{dG2}),
so that the tree-term $\Pi_{tree}=m^2$ just cancels the mass in the
dressed propagator $\Delta$ of Eq.~(\ref{DTL}), giving back the standard free-particle propagator
of Eq.~(\ref{D0}).

Summing up the loops, the transverse dressed propagator
can be written as
\BE
{\Delta}(p)=\left[-p^2-\Pi_{\text{loop}}(p)\right]^{-1}\,,
\label{dressD}
\EE
where $\Pi_{\text{loop}}(p)=\Pi(p)-m^{2}$ is the sum of the transverse part of all the polarization graphs containing
loops (that is, excluding the tree-level term).

The diverging integrals are made finite by dimensional regularization and can be evaluated in the Euclidean space, by setting $d=4-2\epsilon$.
An important feature of the massive expansion is that the crossed graphs cancel all the spurious diverging mass terms exactly, so
that no mass renormalization is required. 
At one-loop, as shown in Refs.~\cite{ptqcd,ptqcd2,xigauge}, in the $\overline{\text{MS}}$ scheme, the diverging part of the proper transverse polarization 
can be written as
\BE
{\Pi}^\epsilon_{\text{loop}} (p)
=-p^2\frac{Ng^2}{(4 \pi)^2}\,\frac{1}{\epsilon}\,\left(\frac{13}{6}-\frac{\xi}{2}\right)\,,
\label{PiTeps}
\EE
which is the same identical result of standard perturbation theory \cite{peskin}.

As usual, the diverging part can be canceled by wave function renormalization, by subtraction
at an arbitrary point. Of course, a finite term $\sim{\rm const.}\times p^2$ arises from the subtraction
and cannot be determined in any way: it depends on the regularization
scheme and on the arbitrary renormalization scale $\mu$, so that its actual value remains somehow arbitrary. When the coupling is absorbed into an overall finite multiplicative renormalization constant for the propagator and the gluon mass $m$ is used to fix the energy scale of the theory (see \cite{ptqcd2} for details), such a term is left as the only spurious free parameter of the approximation.

The fixed-scale approach, as opposed to its RG-improved counterpart \cite{RG}, has the advantage of providing analytical expressions which are
in excellent agreement with the lattice data below 2~GeV -- see Fig.~1 (solid black curve) -- and can be easily
continued to the whole complex plane~\cite{analyt,xigauge,ghost,damp}. The success of the method would suggest that, in the infrared, a mere constant, 
inserted in the factor of $p^2$ in Eq.~(\ref{PiTeps}), can mimic the effects of higher-order terms.

The agreement can be achieved from first principles by using a sort of optimization by variation of the renormalization scheme, a method
that was proven to be very effective for the convergence of the expansion~\cite{stevensonRS,stevensonBook}.
In this framework, the ${\rm const.}\times p^2$ term -- or, equivalently in MOM-like schemes, the dimensionless subtraction scale $\mu/m$~--, is fixed by requiring that some properties related to gauge invariance, and more precisely the gauge invariance of the poles and phases of the residues, are satisfied by the approximate one-loop expression of
the gluon propagator~\cite{xigauge,beta}. 
When such properties are enforced~\cite{xigauge}, the optimized
one-loop analytical expression again provides an excellent agreement with the lattice, as shown in Fig.~1 (dashed green curve). 

\begin{figure}[t] \label{fig:Delta}
\centering
\includegraphics[width=0.35\textwidth,angle=-90]{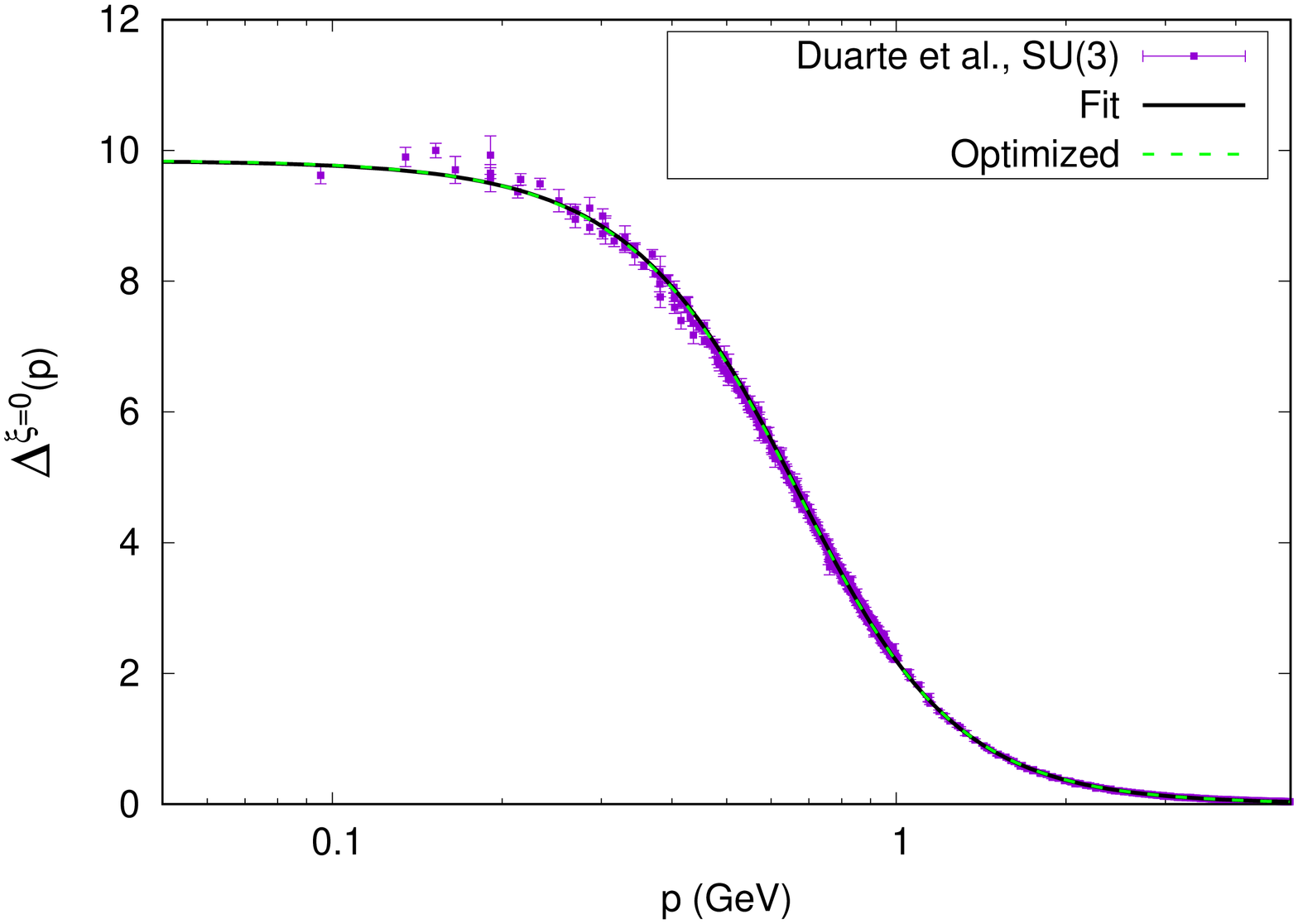}
\caption{The one-loop gluon propagator computed in the Landau gauge ($\xi=0$) within the framework of the fixed-scale screened massive expansion, together with the lattice data of Ref.~\cite{duarte}. Free fit (solid black curve) and optimized calculation (green dashed curve). The energy scale is set by taking $m\approx0.65$~GeV.}
\end{figure}

The gauge invariance of the gluon poles~\cite{kobes90} is an exact property which easily follows from the BRST invariance of the Yang-Mills Lagrangian. Using the BRST symmetry, a Nielsen identity~\cite{nielsen}
\BE
\frac{\partial}{\partial \xi} \frac{1}{\Delta(p)}= 2{\cal F}(p)\,\left[\frac{1}{\Delta(p)}\right]^2\,,
\label{nielsen0}
\EE
where ${\cal F}(p)$ is the transverse component of another Green function (more on this in the next section), can be derived~\cite{breck}. Since the Yang-Mills Lagrangian is not modified as a whole by our shift of the expansion point, we know that, provided that the BRST symmetry is not broken non-perturbatively, a sufficiently accurate approximation of the gluon propagator must have gauge-invariant poles.

Nonetheless, due to the soft breaking of BRST symmetry caused by the introduction of a mass term in the kinetic and interaction terms of the Lagrangian, the gluon propagator computed in the screened expansion does not automatically fulfill such a constraint, for general values of the free parameters. The poles of the gluon propagator were found to be complex~\cite{ptqcd2}, coming in a complex-conjugate pair at $p^{2}=p_{0}^{2},(p_{0}^{2})^{\star}$, with $\text{Im}(p_{0}^{2})\neq 0$. While the poles, being the solution of the $\xi$-dependent equation $\Delta^{-1}(p,\xi)=0$, can in general depend on the gauge parameter $\xi$, the free parameters of the expansion -- that is, the constant in the $\text{const.}\times p^{2}$ term and the gluon mass parameter $m^{2}$ itself -- can be tuned with the gauge so as to make the poles $\xi$-independent~\cite{xigauge}, thus complying with the Nielsen identities. The additional requirement that the \textit{phase} of the residues be gauge-invariant was found to be sufficient to fix the value of the spurious free constant in the Landau gauge~\cite{xigauge} and yielded a gluon propagator which is remarkably close to the free fit obtained from the lattice data, and to the lattice results themselves.

While the gauge invariance of the poles is a trivial consequence of the Nielsen identities, and must hold even when
the poles are complex, the gauge invariance of the phase of the residues is less obvious.
As discussed in Ref.~\cite{xigauge}, because of the square on the right hand side of Eq.~(\ref{nielsen0}), the phase 
of the complex residues is invariant {\it if} the Green function ${\cal F}$ does not have a pole at the same
position as the gluon propagator $\Delta(p)$.

Thus, the discovery of Ref.~\cite{xigauge}, that the gauge invariance of the phase of residues (in the complex plane)
provides an optimal agreement with the lattice data (in the Euclidean space), might lead to two different interpretations:
either the function ${\cal F}$ has no poles at the same position as the gluon propagator, and the gauge invariance of the phase of the gluon residues is an exact property, or the change of the phase with the gauge is an accidentally small higher-order effect which is not seen at one-loop. Actually, due to the arbitrariness in the renormalization of the gluon residue, there is also a third possibility, which will be discussed in the next section.

A more detailed analysis of the point motivates the explicit derivation of the Nielsen identity, Eq.~(\ref{nielsen0}), in the
special context of the screened expansion. This will be the content of the next two sections.

\section{The Nielsen identities}

The Nielsen identities~\cite{nielsen} are a set of equations which determine the gauge dependence of the exact Green functions of a gauge theory. They can be derived using BRST symmetry, under the hypothesis that the latter is not broken in the vacuum. For the propagators of QCD, the identities were fully discussed and derived 
by a functional method in Ref.~\cite{breck}. In that work, a detailed calculation was
reported for the explicit one-loop identities, in the framework of standard perturbation theory.

In this Section, we give a quite straightforward derivation of the identities and discuss their relation with the polarization function. A detailed derivation of the explicit one-loop expressions, in the framework of the screened massive expansion, will be discussed in the next section.\\

We start by re-introducing the Nakanishi-Lautrup auxiliary field $B_{a}$ in the Yang-Mills action and writing the gauge-fixing term in Eq.~(\ref{fix}) as
\BE
{\cal L}_{fix}=\frac{\xi}{2} B_aB_a+ B_a(\partial\cdot A_a)\,.
\label{fixB}
\EE
Integrating out the Nakanishi-Lautrup field is equivalent to solving the equation of motion $B_a=-(\partial\cdot A_a)/\xi$. 

When expressed in terms of the $B$-field, the total Yang-Mills Lagrangian satisfies the usual BRST invariance property
\BE
\delta_\theta {\cal L}=0\,,
\EE
where 
\begin{align}
\delta_\theta A^\mu_a&=\theta D^\mu c_a\,,\nn\\
\delta_\theta \bar c_a&=\theta B_a\,,\nn\\
\delta_\theta c_a&=-\frac{g}{2} f^{abc} \theta c_b c_c\,,\nn\\
\delta_\theta B_a&=0\,,
\label{BRST}
\end{align}
and $\theta$ is a Grassmann parameter.
 
The field $B$ also determines the gauge-parameter dependence of the total Lagrangian, since the derivative of the latter with respect to $\xi$ is given by
\BE
\frac{\partial {\cal L}}{\partial\xi}=\frac{1}{2} B_a B_a\,.
\label{dL}
\EE
Using Eq.~\eqref{dL}, the average $\langle {\cal O}\rangle$ of any operator ${\cal O}$,
\BE
\langle {\cal O}\rangle=\frac{\int {\cal O} \exp\left(i\int {\cal L}\right)}{ \int \exp\left(i\int {\cal L}\right)}\,,
\EE
is easily seen to satisfy
\begin{align}
\frac{\partial}{\partial \xi} \langle {\cal O}\rangle&= \langle {\cal O}\left(i\int\frac{\partial {\cal L}}{\partial\xi}\right)\rangle
- \langle {\cal O}\rangle\, \langle \left(i\int\frac{\partial {\cal L}}{\partial\xi}\right)\rangle=\nn\\
&=\frac{i}{2}\langle {\cal O} \int B_aB_a\rangle\,.
\end{align}
The last equality follows from Eq.~(\ref{dL}) and from the Slavnov-Taylor identity
\BE
0= \langle \delta_\theta(\bar c_aB_a)\rangle=\langle \delta_\theta\bar c_a B_a\rangle=\langle B_aB_a\rangle\,,
\label{BB}
\EE
which holds provided that the vacuum is BRST-invariant, so that
\BE
\langle\delta_{\theta}\mathcal{O}'\rangle=0
\EE
for any operator $\mathcal{O}'$.

For the case of the exact gluon propagator 
\BE
\Delta^{\mu\nu}_{ab}(x,y)=-i\langle A^\mu_a(x)A^\nu_b(y)\rangle\,,
\label{Dg}
\EE
we can denote by $2{\cal F}$ the Green function
\begin{align}
2{\cal F}^{\mu\nu}_{ab}(x,y)&=
\frac{\partial}{\partial \xi} \Delta^{\mu\nu}_{ab}(x,y)\nn\\
&=\frac{1}{2}\int {\rm d}^4z \langle A^\mu_a(x)A^\nu_b(y) B_c(z)B_c(z)\rangle\,,
\label{prenielsen}
\end{align}
or in a more compact notation
\BE
\frac{\partial\Delta}{\partial \xi}=\frac{1}{2}\langle AA\int B^2\rangle=2{\cal F}\,.
\EE
Denoting by $\Gamma=-\Delta^{-1}$ the two-point vertex function,
the Nielsen identity for $\Gamma$ reads
\BE
\frac{\partial\Gamma}{\partial \xi}=\Delta^{-1}\cdot\frac{\partial\Delta}{\partial \xi}\cdot\Delta^{-1}=\Gamma\cdot (2{\cal F}) \cdot\Gamma\,,
\label{nielsen1}
\EE
where the dot products are functional products which become ordinary products when the Fourier transform is taken.

Strictly speaking, Eq.~(\ref{nielsen1}) becomes the Nielsen identity only when an independent and direct evaluation of the Green function ${\cal F}$ is provided by the Slavnov-Taylor identities, as shown below.
While the Green function ${\cal F}$ seems to have two gluon legs (hence a double gluon pole in the Fourier transform) in Eq.~(\ref{prenielsen}), one of the legs is eaten up as a result of BRST symmetry, so that according to Eq.~(\ref{nielsen1}) the Fourier transform of the function ${\cal F}$ has a single pole at most.
This follows from the Slavnov-Taylor identity
\BE
0=\langle \delta_\theta(AA\bar cB)\rangle=\langle (Dc)A\bar cB\rangle+\langle A(Dc)\bar cB\rangle+\langle AABB\rangle\,,
\EE
which yields
\begin{align}
2{\cal F}^{\mu\nu}_{ab}(x,y)=-\frac{1}{2}\bigg[
&\int {\rm d}^4z\>\langle D^\mu c_a(x)A^\nu_b(y) \bar c_c(z)B_c(z)\rangle\nn\\
&+ \> (x\leftrightarrow y,a\leftrightarrow b)\>\>\bigg]\,.
\label{nielsen2}
\end{align}
The equivalence of the function ${\cal F}$ in Eq.~(\ref{nielsen2}) and in Eq.~(\ref{prenielsen}) is the core of the Nielsen
identity for the gluon propagator.

The presence of a single gluon leg in Eq.~(\ref{nielsen2}) ensures that at least one of the two $\Gamma$ factors survives on the right hand side
of Eq.~(\ref{nielsen1}) and
that a zero occurs at the pole position $p=p_0$ in the derivative of the Fourier transforms:
\BE
\left[\frac{\partial \Gamma}{\partial \xi}\right]_{p=p_0}=\left[\frac{\partial \Pi}{\partial \xi}\right]_{p=p_0}=0 \quad {\rm if}
\quad \Gamma(p_0)=0\,.
\label{Pzero}
\EE
As a consequence, the position of the pole is gauge-parameter-independent in the exact gluon propagator, as can be explicitly seen from the equations
\begin{align}
0 &= \frac{d}{d\xi} \Gamma(p_{0}^{2}(\xi),\xi)=\nn\\
&=\frac{\partial \Gamma}{\partial \xi}(p_{0}^{2}(\xi),\xi)+\frac{\partial \Gamma}{\partial p^{2}}(p_{0}^{2}(\xi),\xi)\,\frac{dp_{0}^{2}}{d\xi}(\xi)=\nn\\
&=\frac{\partial \Gamma}{\partial p^{2}}(p_{0}^{2}(\xi),\xi)\,\frac{dp_{0}^{2}}{d\xi}(\xi)\quad\Longrightarrow\quad\frac{dp_{0}^{2}}{d\xi}=0\,,
\end{align}
given that in order for $\Gamma$ to have a zero at $p_{0}^{2}$, $(\partial\Gamma/\partial p^{2})_{p_{0}}$ must be finite. Here, a transverse projection is understood for all the functions, since the longitudinal parts are known exactly
and, in that case, the invariance of the pole is trivial, being the longitudinal pole unshifted from $p=0$ in any gauge. 

The discovery that the phase of the complex residue may be gauge-invariant~\cite{xigauge} has led to the claim that the Green function ${\cal F}$ might have no pole at all in $p=p_0$. Then, the double zero on the right hand side of Eq.~(\ref{nielsen1}) due to the $\Gamma$'s would be enough for ensuring that
\BE
\frac{\partial}{\partial \xi} \left[\frac{\partial \Gamma}{\partial p^2}\right]_{p=p_0}=0\,,
\EE
yielding a proof of gauge invariance for the residue~\cite{xigauge}, see ahead. Going back to Eq.~(\ref{nielsen1}), the claim is equivalent to assuming that the derivatives ${\partial \Gamma}/{\partial \xi}$ and ${\partial \Pi}/{\partial \xi}$ have a double zero at the pole position.

Actually, the invariance of the {\it modulus} of the residue would not make much physical sense, since the modulus is defined up to an arbitrary -- potentially gauge-dependent -- real renormalization factor. What emerged in Ref.~\cite{xigauge} was the invariance of the {\it phase} of the residue, which would be enforced by
a weaker condition: denoting by $R=\vert R\vert \exp(i\theta)$ the complex residue at the pole $p^2=p^2_0$, the
transverse projection of the  two-point function
reads
\BE
\Gamma(p^2)=(p^2-p_0^2)\frac{e^{-i\theta}}{\vert R\vert}+\dots\,,
\label{zero}
\EE
and because of the gauge invariance of the pole $p_0$, the logarithmic derivative gives
\BE
\frac{\partial \theta}{\partial \xi}=-\Im\left[\frac{1}{\Gamma}\frac{\partial \Gamma}{\partial\xi}\right]_{p=p_0}=
-\Im\left[(2{\cal F})\,\Gamma\right]_{p=p_0}\,,
\label{imag}
\EE
where the second equality follows
from Eq.~(\ref{nielsen1}) and a transverse projection is understood in all the functions on the right hand side.
Thus, the vanishing of the imaginary part, on the right hand side, would be enough for ensuring that the phase is invariant.

While the invariance of the modulus of the residue seems to be unnecessary in view of renormalization, the invariance of the phase makes sense in presence of complex poles: the phase determines the shape of the principal part of the propagator -- which, as shown in \cite{xigauge}, makes up for the largest contribution to $\Delta(p)$ -- and has an effect on all related observable objects, like the Fourier transform of the propagator which at large momenta could be seen as a short-distance approximation for the quark-quark potential. On the other hand, 
the invariance of the phase has always been expected whenever the pole and residue were real, since then, trivially,  $\theta=0$ for any $\xi$.

It is instructive to explore the content of Eq.~(\ref{prenielsen}) in terms of diagrams of the screened expansion and, more generally, of perturbation theory.
We will be interested in the Fourier transform
\BE
2\delta_{ab}{\cal F}^{\mu\nu}(p)\equiv 2{\cal F}^{\mu\nu}_{ab}(p,-p)\,,
\EE
where
\begin{align}
&2{\cal F}^{\mu\nu}_{ab}(p,q)\,(2\pi)^{4}\,\delta^{(4)}(p+q)=\nn\\
&=2\int  {\rm d}^4x \,{\rm d}^4y \, {\cal F}^{\mu\nu}_{ab}(x,y) e^{ip\cdot x+iq\cdot y}\,.
\label{FTF}
\end{align}

We first recover, by the equations of motion,
\BE
\langle B_a A^\mu_b\rangle=-\frac{1}{\xi} \langle (\partial\cdot A_a) A^\mu_b\rangle\,,
\EE
yielding the exact result 
\BE
\int  {\rm d}^4x e^{ip\cdot x} \langle B_a(0) A^\mu_b(x)\rangle=\frac{-ip_\nu}{\xi}i\Delta^{\nu\mu}_{ab}(p)\
=\frac{\delta_{ab}p^\mu}{-p^2}\,,
\label{BA}
\EE
which is valid to all orders~\cite{breck} because of Eq.~(\ref{DTL}). All graphs contributing to the right hand side of Eq.~(\ref{prenielsen}) can be obtained by the insertion of the two-point local vertex $B_a^2$
in the graphs of the gluon propagator. At tree level, there is only one term given by the product 
$\langle AB\rangle\langle BA\rangle$ with a symmetry factor of 2, yielding
\BE
2{\cal F}^{\mu\nu}(p)=\left[\frac{p^\mu}{-p^2}\right]\left[\frac{-p^\nu}{-p^2}\right]=-\frac{p^\mu p^\nu}{p^4}\,.
\label{tree}
\EE
Inserting this longitudinal term in Eq.~(\ref{nielsen1}), together with the exact longitudinal part
$\Gamma^{\mu\nu}=\ell^{\mu\nu}p^2/\xi$, the identity is easily seen to be satisfied exactly. Thus, we predict that all loop contributions
to ${\cal F}$ must be transverse~\cite{breck}. In fact, this is the case, since they can all be derived by  insertion of a vertex $\int B^2$ in all  graphs for the gluon polarization, which is transverse.

In more detail, denoting by $\pi^{(n)}$ a polarization graph with $n$ internal gluon lines, all the corresponding graphs for
the function ${\cal F}$, in Eq.~(\ref{prenielsen}), are obtained by substituting the longitudinal term $\langle AB\rangle\langle BA\rangle$
for one of the $n$ internal gluon lines and restoring the external gluon legs. The inserted longitudinal term is just the tree-level graph for $2{\cal F}$ and is equal to $\Delta^L\ell^{\mu\nu}/\xi=\partial \Delta_m^{\mu\nu}/\partial \xi$ 
according to Eqs.~(\ref{Deltam}), (\ref{DTL}) and (\ref{tree}). Thus,
we are just replacing a gluon line by its longitudinal part, divided by $\xi$. Now, this is precisely what we get by taking
the derivative $\partial \pi^{(n)}/\partial \xi$. Summing over $n$, we get a direct proof of Eq.~(\ref{prenielsen}), 
since the dependence on $\xi$ is in the internal lines, while
the external legs are projected on the transverse polarization and do not depend on $\xi$. This argument holds both
for the screened expansion ($m\not=0$) and for standard perturbation theory ($m=0$), since it only depends on the transversality of the exact gluon polarization. 

While the content of Eq.~(\ref{prenielsen}) is trivial in terms of diagrams, its equivalence to  Eq.~(\ref{nielsen2}) is not
immediate and there is no one-to-one correspondence of diagrams.  Eq.~(\ref{nielsen2}) follows from
the overall BRST symmetry of the Lagrangian and the equivalence to Eq.~(\ref{prenielsen}) holds for the exact
functions. 
As shown in Ref.~\cite{breck}, in ordinary perturbation theory, 
if both functions are correctly expanded in powers of the coupling, they must agree at any finite order. On the other hand, the addition and subtraction of a gluon mass term that defines the screened massive expansion causes the soft breaking of BRST invariance at any finite order. Thus, we expect the Nielsen identities not to hold perturbatively in the screened expansion. This does not imply, however, that the screened expansion's gluon poles are not gauge-invariant. As discussed in the last section, the freedom in the choice of the spurious free parameters is still enough to enforce their invariance, once an explicit expression for the gluon propagator has been obtained at finite order.

At any finite order, deviations from the exact BRST symmetry are a {\it measure} of the
accuracy of the truncated expansion~\cite{xigauge,ghost,beta}. Thus, it is not a case that the screened expansion gives an excellent agreement with the lattice data when optimized by the constraints of pole (and
phase) invariance, since these are the conditions which minimize the deviations between one-loop and exact results.

On the other hand, we might wonder if the gauge-invariance of the phase of the residue is an exact property of Yang-Mills theory. To date, we have not been able to reach a formal proof. For what concerns the screened expansion, due to the previously discussed soft breaking of BRST invariance, the perturbative expression for $\partial\theta/\partial \xi$ cannot be trusted as is at any finite order. Nonetheless, a non-perturbative resummation of the gluon graphs in the Nielsen identity might give us hints as to whether the phase is really invariant. This will be discussed in Sec.~V.

\section{One loop explicit calculation}

The Green function ${\cal F}$, in Eq.~(\ref{nielsen2}), can be directly evaluated in the framework of the 
screened massive expansion, order by order. Here, we give the explicit result up to one-loop. 

At tree level, there is only one graph contributing to Eq.~(\ref{nielsen2}) which factors as
$\langle (\partial^\mu c) \bar c\,\rangle\langle BA^\nu\rangle$ yielding
\BE
2{\cal F}^{\mu\nu}(p)=-(-ip^\mu)\left(\frac{i}{p^2}\right)\left(\frac{-p^\nu}{-p^2}\right)=-\frac{p^\mu p^\nu}{p^4}
\label{tree2}
\EE
in agreement with the exact result in Eq.~(\ref{tree}). There are no crossed graphs at tree level because any insertion of the transverse mass vanishes in the longitudinal tree term.  As discussed above, since Eq.~(\ref{tree2}) gives the whole
longitudinal contribution, the sum of all higher order terms must be transverse.
In fact, the uncrossed one-loop graphs which contribute to Eq.~(\ref{nielsen2}) are the first three pairs reported in Fig.~2,
and each pair gives a pure transversal term.
More generally, all loop graphs occur in pairs, with the structure displayed in Fig.~3,
arising from the splitting of the covariant derivative in two terms,
\BE
D^\mu c_a=\delta_{ab}\partial^\mu c_b+gf_{abc} A_b^\mu c_c\,,
\EE
and from the insertion of a ghost-gluon vertex  $(\bar c A c)$ in the first of these.
The expressions of the graphs of type (1) and
(2) have the following general form, respectively,
\begin{align}
2{\cal F}^{\mu\nu}_{ab\,(1)}(p)&=-i\langle (\partial^\mu c_a) \, (gf_{def}\partial_{\alpha}\bar c_d A^\alpha_e c_f)\cdots\rangle=\nn\\
&=(-ip^\mu)\left(\frac{1}{p^2}\right) (ip_\alpha) g f_{aef}\int {\Delta_m}^{\alpha\beta}_{\,eg}\cdots\,,\nn\\
2{\cal F}^{\mu\nu}_{ab\,(2)}(p)&=-\langle (g f_{aef} A^\mu_e c_f) \cdots\rangle=\nn\\
&=-g f_{aef}\int {\Delta_m}^{\mu\beta}_{\,eg}\cdots\,,
\label{transvpair}
\end{align}
and their sum is a manifestly transverse contribution
\BE
2{\cal F}^{\mu\nu}_{ab\,(1+2)}(p)=
\left(g^{\mu}_{\,\alpha}-\frac{p^\mu p_\alpha}{p^2}\right) g f_{eaf}\int {\Delta_m}^{\alpha\beta}_{\,eg}\cdots\,.
\label{trans}
\EE

\begin{figure}[b] \label{fig:graphs1}
\vskip 1cm
\centering
%\hspace*{1cm}
\includegraphics[width=0.25\textwidth,angle=-90]{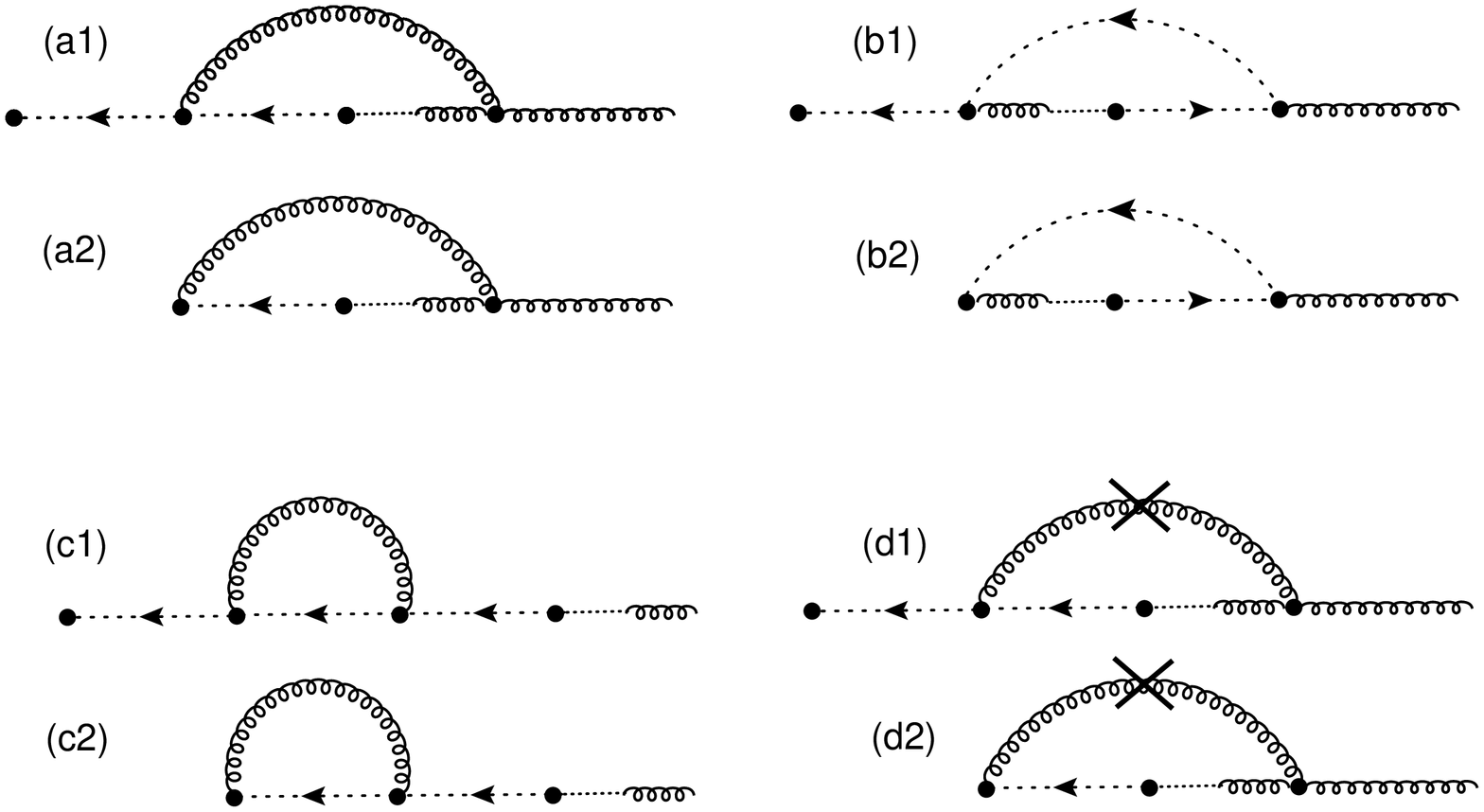}
\caption{One-loop graphs contributing to the function ${\cal F}$, as defined in Eq.(\ref{nielsen2}), in the screened expansion. 
The mixed line is the longitudinal function $\langle BA\rangle$, while the solid cross is the transverse mass counterterm.} 
\end{figure}

\begin{figure}[b] \label{fig:graphs2}
\vskip 1cm
\centering
%\hspace*{1cm}
\includegraphics[width=0.25\textwidth,angle=-90]{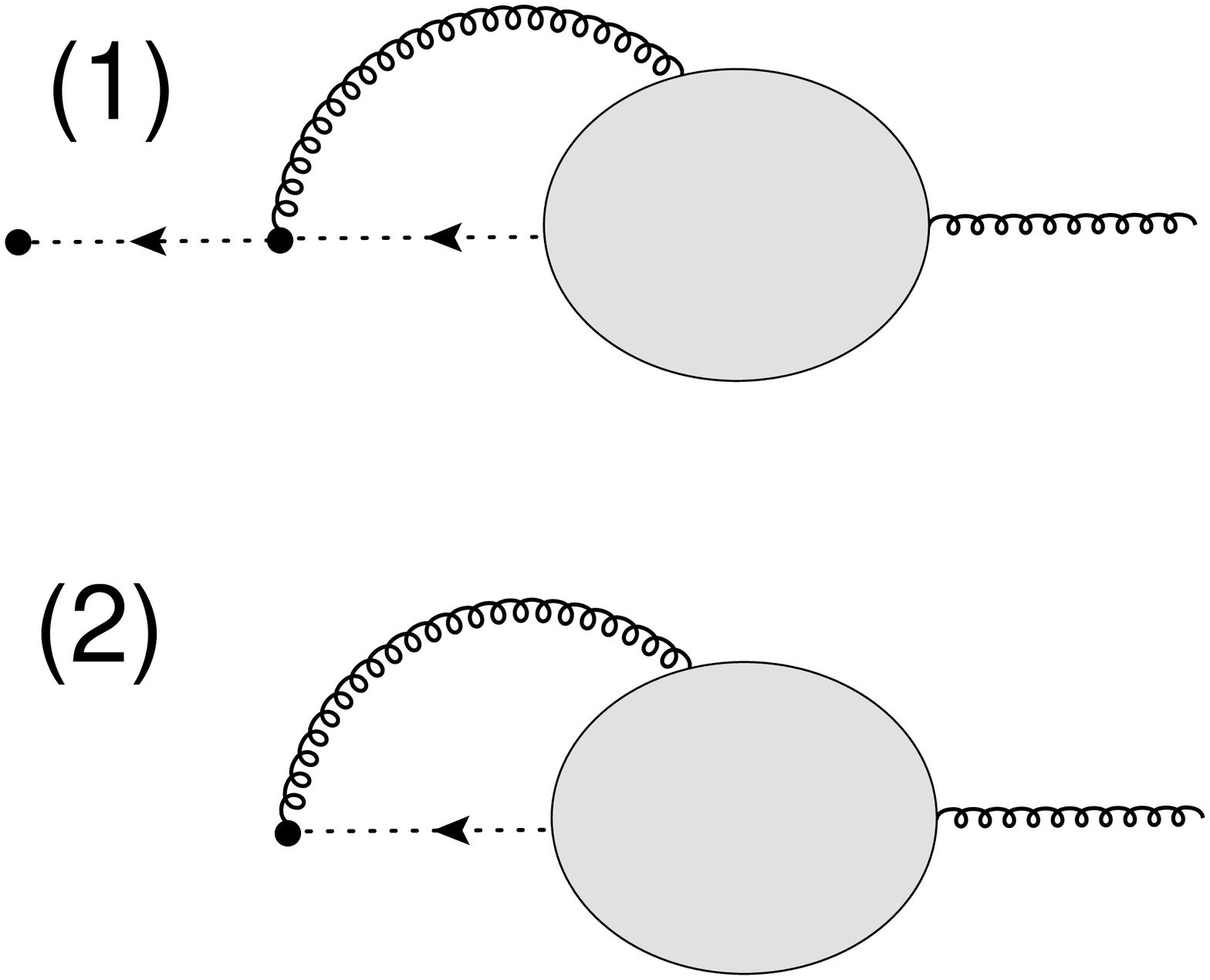}
\caption{General structure of each pair of graphs contributing to the function ${\cal F}$ in Eq.~(\ref{nielsen2}). For each pair, the
sum of graph (1) and graph (2) gives a transversal term, as shown in Eqs.~(\ref{transvpair}) and (\ref{trans}).} 
\end{figure}

As we said, these three pairs give the transverse one-loop contribution for any expansion with
a mass in the free propagator, $\Delta_m$.  While our main interest is on the screened massive expansion of Section II,
their explicit expressions might be of some interest for other theories, like Curci-Ferrari model in the Landau gauge.
We must mention that, in massive theories, there is a class of anomalous graphs, not shown in Fig.~2, contributing to the
longitudinal part of ${\cal F}$. These arise from polarization insertions in the external gluon leg of the 
longitudinal tree-level graph.
While individual polarization terms might have a non vanishing longitudinal part in massive theories, their exact resummation
is zero in the screened massive expansion, ensuring that the tree-level term still provides the total longitudinal 
contribution, as required by the BRST symmetry. Thus, we might neglect the anomalous terms entirely.

The third pair of graphs in Fig.~2, (c1) and (c2), have a longitudinal leg on the right side. Then, their sum is zero according to
Eq.~(\ref{trans}), because of the transverse projector coming from the loops on the left.
The second pair, graphs (b1) and (b2), are basically the same as
in the standard perturbation theory, since no massive propagator occurs in the loop. The only difference arises from the
bare gluon leg on the right side, which must be replaced by the massive free propagator $\Delta_m$ in the screened expansion. 
The first pair, graphs (a1) and (a2), differs from the standard result because of the internal  massive gluon line. The explicit calculation
is straightforward and the detailed steps are reported in the appendix. The sum of all the uncrossed one-loop graphs can be written as
\BE
2{\cal F}^{\mu\nu}(p)=\frac{g^2N}{64\pi^2} \, \frac{t^{\mu\nu}(p)}{ p^2-m^2}\, F(-p^2/m^2)\,,
\label{1loopunc}
\EE
where the diverging function $F(s)$, with $s=-p^2/m^2$, is regularized by setting $d=4-2\epsilon$ and reads
\BE
F(s)=\frac{2}{\epsilon}-3L(s)+\log(s) -2\xi+ {\rm const.}\,,
\label{Func}
\EE
while the logarithmic function $L(s)$, which is derived in Eq.~(\ref{Ls}), can be recast as
\begin{align}
L(s)=&-\frac{1}{3s}+\left(1-\frac{1}{s}+\frac{1}{3s^2}\right)\log s\nn\\
&+\left[ \left( s+1-\frac{1}{s}+\frac{1}{3s^2}\right)\log\left(1+\frac{1}{s}\right)-1\right]
\label{Ls2}
\end{align}
and shows the leading behavior $L(s)\sim \log s$ in the limit $s\to \infty$, which occurs when the mass is set to zero in order to
recover the result of standard perturbation theory. Actually, as shown in the appendix, graphs (a) and (b) agree with the known results
in that limit~\cite{breck}. Moreover, in the same limit, 
the diverging part can be checked by a direct comparison with  the explicit one-loop diverging term of the
polarization, which is well
known and is reported in Eq.~(\ref{PiTeps}). By a direct calculation of the derivative and by inserting it in Eq.~(\ref{nielsen1}), with 
the tree-level
two-point function, $\Gamma(p)=p^2$, the diverging part of the transverse one-loop function ${\cal F}$ reads 
\begin{align}
2{\cal F}=&\frac{1}{\Gamma^2}
\frac{\partial\Gamma}{\partial \xi}
=\frac{1}{\Gamma^2}
\frac{\partial\Pi}{\partial \xi}\nn\\
&\sim\left[\frac{1}{p^2}\right]^2 
\left[ -p^2\frac{Ng^2}{(4 \pi)^2}\left(\frac{-1}{2\epsilon}\right)\right]\nn\\
&=\frac{g^2N}{64\pi^2} \, \frac{1}{ p^2}
\left[ \frac{2}{ \epsilon}\right]\,,
\end{align}
in perfect agreement with Eqs.~(\ref{1loopunc}) and (\ref{Func}). That is important for the renormalization of the function ${\cal F}$, since all the divergences must be absorbed by the wave function renormalization of the gluon propagator in order to make sense of the Nielsen identity, Eq.~({\ref{nielsen1}), when the finite, renormalized propagator is considered.

It is instructive to see how the divergence cancels in the renormalized functions.
In the $\overline{\text{MS}}$ scheme,  the wave function renormalization constant $Z_A$ 
follows from the divergence of the polarization in Eq.~(\ref{PiTeps}) and reads
\BE
Z_A=1+\frac{g^2N}{(4\pi)^2}\left[\frac{1}{\epsilon}\right]\left(\frac{13}{6}-\frac{\xi}{2}\right)\,,
\label{ZA}
\EE
while the logarithmic derivative of the renormalized (transverse) vertex function, $\Gamma^R=Z_A\Gamma$, can be written as
\begin{align}
\frac{1}{\Gamma^R}\frac{\partial\Gamma^R}{\partial \xi}&=\frac{1}{Z_A}\left(\frac{\partial Z_A}{\partial \xi}\right)
+\frac{1}{\Gamma}\frac{\partial\Gamma}{\partial \xi}\nn\\
&=-\frac{g^2N}{2(4\pi)^2}\left[\frac{1}{\epsilon}\right]+\frac{g^2N}{64\pi^2}\,F(s)\nn\\
&=\frac{g^2N}{64\pi^2}\left[ F(s)-\frac{2}{\epsilon}\right]\,,
\end{align}
where the second  term in the second line arises from the Nielsen identity, Eq.~(\ref{nielsen1}), and from the insertion of the one-loop result, Eq.~(\ref{1loopunc}), neglecting higher order terms. 
According to Eqs.~(\ref{zero}) and (\ref{imag}), the real and imaginary part of the logarithmic derivative are the gauge-parameter 
derivative of the modulus and phase, respectively, of the residue. They are made finite by the subtraction of the diverging term which occurred in Eq.~(\ref{Func}). But, while the modulus still depends on an arbitrary (real) constant which arises from the subtraction and
regularization schemes, as it should, the phase of the residue is finite anyway and does not depend on the renormalization
up to higher order corrections. In fact, we can write
\BE
\frac{\partial \theta}{\partial \xi}=-\Im\left[\frac{1}{\Gamma}\frac{\partial \Gamma}{\partial\xi}\right]_{p_0}=
\left(\frac{g^2N}{64\pi^2}\right)\,\Im\left[\,F\,\Gamma\,\Delta_m\,\right]_{p_0}\,,
\label{imag2}
\EE
where $p_0$ is the pole position and, neglecting higher order corrections, $\Gamma$ can be taken at tree level, so that
$\Gamma\Delta_m\approx -1$, which is real. Thus, when computing the derivative of the phase, we can drop
all constants and real additive terms in the one-loop function $F$, which simplifies as
\BE
F(s)\to \log(s)-3L(s)\,.
\label{Fs2}
\EE

At one loop, in principle, there are other graphs contributing to $F$ in the screened expansion: the crossed graphs which contain one or more insertions of the transverse mass
counterterm, such as diagrams (d1) and (d2) in Fig.~2. These are higher order graphs by vertex counting, but still one-loop if the powers of $g^2$ are considered. 
Thus, their inclusion must be discussed in the framework of the detailed approximation scheme which is used.
For instance, the inclusion of an infinite set of graphs with any number of mass counterterms is equivalent to a Dyson resummation of the constant polarization
term $\Pi=m^2$. The effect is that, in any gluon line,  the massive gluon propagator is replaced by the bare massless one, restoring 
the ordinary standard perturbation theory.   That is not what we would aim to, of course. The inclusion of a finite number of
mass counterterms, up to a given order, turned out to be the best choice for canceling the spurious divergences without falling
into a trivial resummation~\cite{ptqcd2}. Here, no spurious divergence is found and the inclusion of a finite number of crossed
graphs will be discussed case by case. 

The crossed graphs can be easily evaluated by derivatives of the standard one-loop graphs~\cite{ptqcd,ptqcd2,analyt}.
For instance, the fourth pair of graphs in Fig.~2,  graphs (d1) and (d2),  contain  one insertion of the transverse mass counterterm
in the internal gluon line which is replaced by the transverse chain $\Delta_m\cdot m^2\cdot  \Delta_m$
\begin{align}
\frac{1}{-p^2+m^2} \to & \frac{1}{-p^2+m^2}\, m^2\,\frac{1}{-p^2+m^2}\nn\\
&=-m^2\frac{\partial}{\partial\, m^2} \left[\frac{1}{-p^2+m^2}\right]\,.
\end{align}
Thus, after amputating the external leg, the inclusion of the crossed graphs (d1) and (d2) follows from the corresponding uncrossed graphs, (a1) and (a2), as
\BE
\Gamma_m{\cal F}_{(a)} +\Gamma_m{\cal F}_{(d)} 
= \left(1-m^2\frac{\partial}{\partial\, m^2} \right) \left[\Gamma_m {\cal F}_{(a)}\right]\,.
\EE
That is equivalent to replacing the logarithmic function $L(s)$ with a new function $L^C(s)$ in Eqs.~(\ref{Func})
and (\ref{Fs2}),
defined as
\BE
L^C(s)=\left(1+s\frac{\partial}{\partial s}\right) L(s)-1\,,
\EE
where the added constant, $-1$, arises from the derivative of  $1/\hat \epsilon$ according to Eq.~(\ref{eps}).
The explicit calculation yields
\begin{align}
L^C(s)=&\frac{1}{3s}-\frac{4}{3(1+s)}
+\left(1-\frac{1}{3s^2}\right)\log s\nn\\
&+\left[ \left(2s+1-\frac{1}{3s^2}\right)\log\left(1+\frac{1}{s}\right)-2\right]\,.
\label{LC}
\end{align}
The manifest leading behavior $L^C(s)\sim L(s)\sim \log s$, in the limit $s\to \infty$, confirms that the
contribution of the crossed graphs is zero in the limit $m\to 0$.

\section{Discussion}

We would like to discuss the invariance properties of the principal part of the gluon propagator. Our starting point
is Eq.~(\ref{imag2}), which gives the explicit one-loop gauge dependence for the phase of the residue at the pole
of the gluon propagator, as computed in the screened expansion by the Nielsen identities. With $\Delta_{m}\Gamma=-1$ at tree level, the identity simplifies as
\BE
\frac{\partial \theta}{\partial \xi}=-\Im\left[\frac{1}{\Gamma}\frac{\partial \Gamma}{\partial\xi}\right]_{p_0}=
-\left(\frac{g^2N}{64\pi^2}\right)\,\Im\left[\,F\,\right]_{p_0}\,.
\label{imag3}
\EE

As previously discussed, since BRST invariance is broken in the screened expansion at any finite order, we may expect $\Im[F]_{p_{0}}$ to be different from zero even if the phase were exactly gauge-invariant. This is indeed the case, as we show in Fig.~4 by plotting the $\Im \{F(-p^2/m^2)\}=0$ contour in the complex $p$-plane. We find a continuous line of zeros for the imaginary part, but quite far away from the pole position (asterisk-shaped point in the figure), which was found at $p_0/m=0.8857+0.5718i$ in Ref.~\cite{xigauge} by the optimized one-loop expansion. The line does not cross the pole, but we would not expect that to happen at one-loop.

\begin{figure}[t] \label{fig:contour1}
\centering
\includegraphics[width=0.35\textwidth,angle=-90]{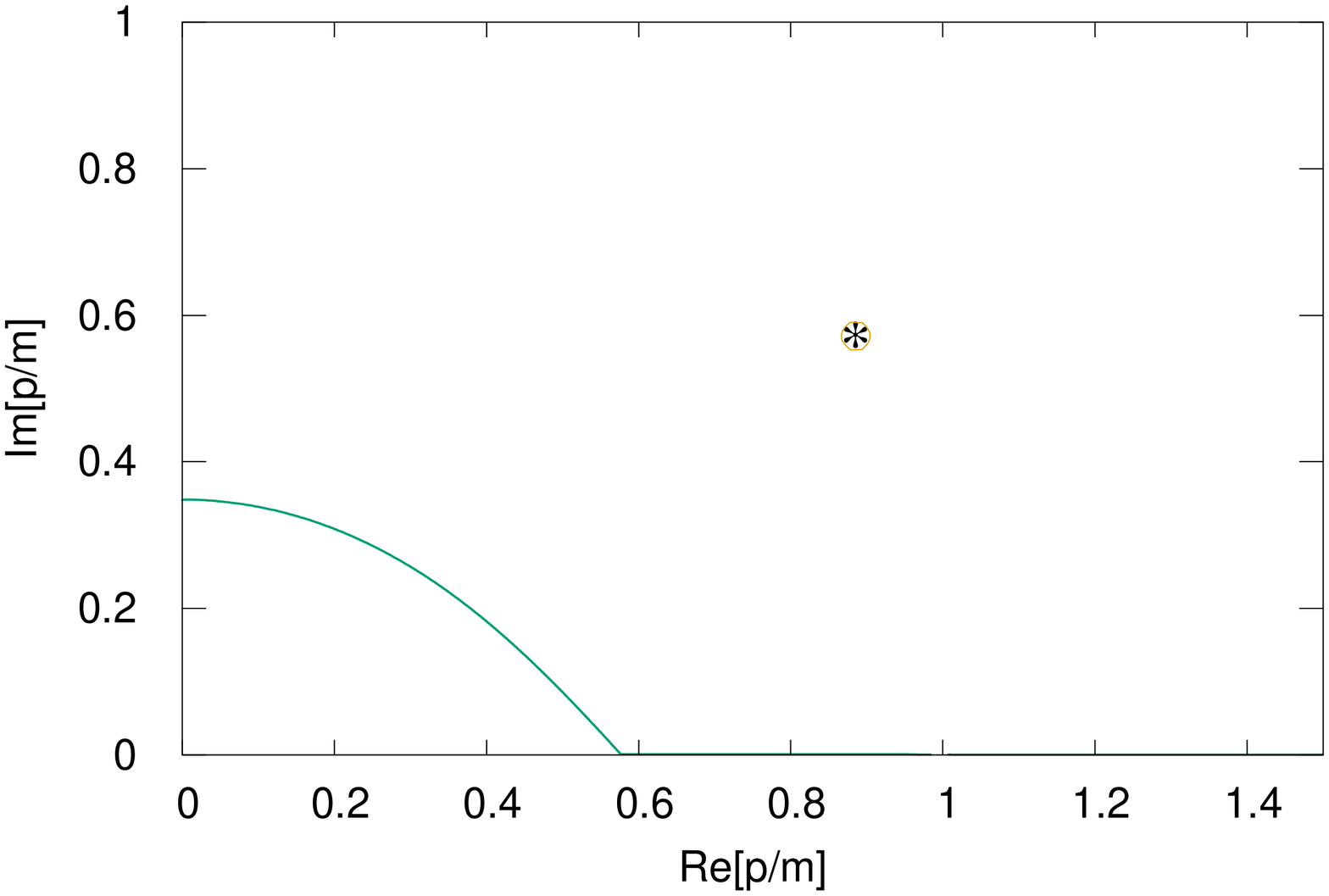}
\caption{Contour plot of the equation $\Im\{F(-p^2/m^2)\}=0$ in the complex plane of $p$, in units of $m$. The asterisk
is the gluon pole $p_{0}$, $\Rerm{[p_0/m]}=0.8857$, $\Im{[p_0/m]}=0.5718$, as found in Ref.~\cite{xigauge}. A continuous line of zeros is
found for the imaginary part of $F$. The plain Eq.~(\ref{Fs2}) is used for the calculation, without any insertion of crossed graphs or resummations.}
\end{figure}

Inserting a finite number of mass counterterms in the internal gluon lines would not change the result too much. For instance, by including the crossed graphs (d1) and (d2) we obtain, by Eq.~(\ref{LC}), the contour plot shown in Fig.~5. Again, we find a line of zeros, but the distance from the pole is even larger.

\begin{figure}[t] \label{fig:contour2}
\centering
\includegraphics[width=0.35\textwidth,angle=-90]{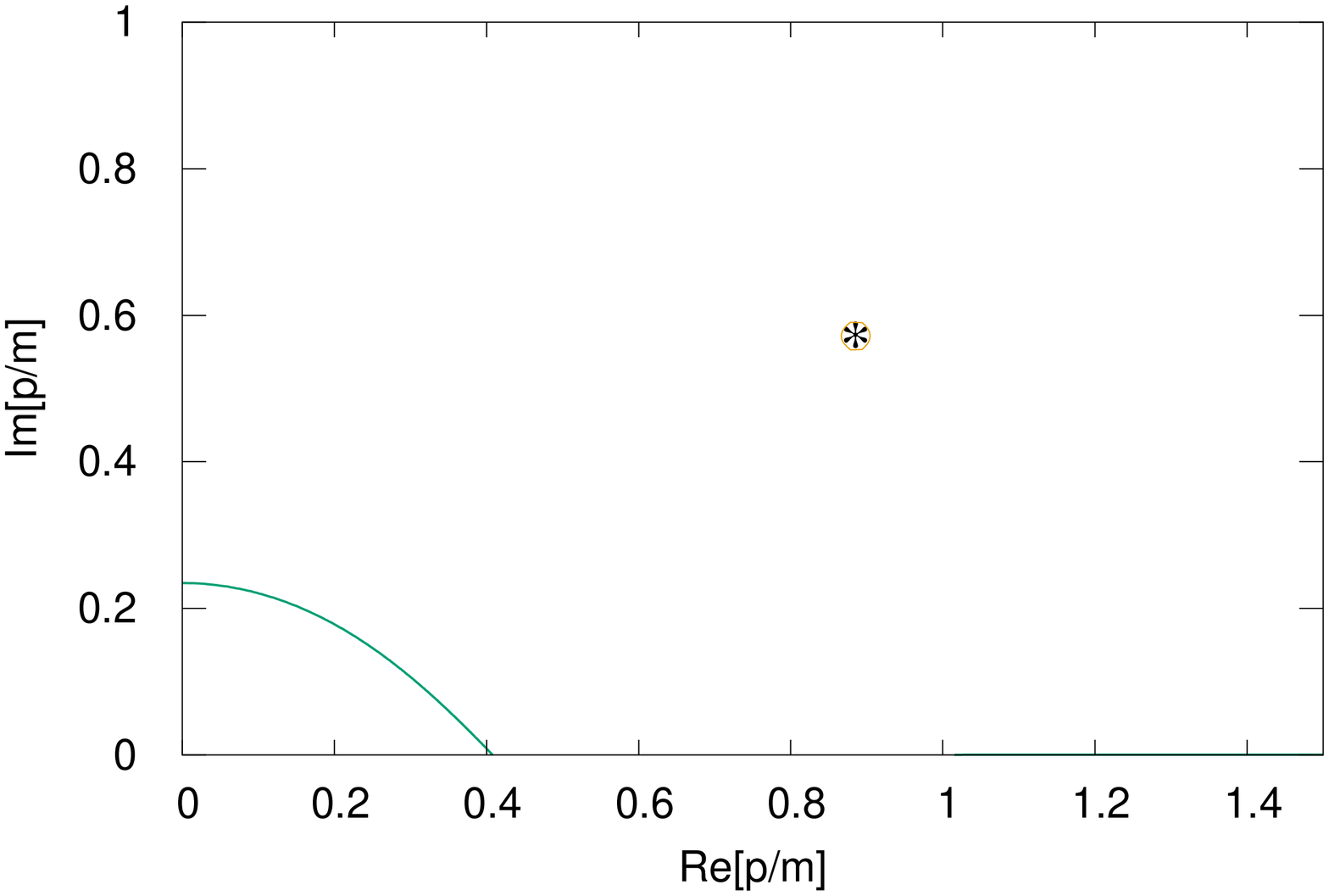}
\caption{Same as Fig.~4,  but with the inclusion of the crossed graphs (d1) and (d2). Eq.~(\ref{LC}) is used for the calculation instead of Eq.~(\ref{Ls2}).}
\end{figure}

A more accurate approximation of the exact result would consist in the resummation of all the one-loop polarization insertions in the internal gluon line. This would be equivalent to replacing the bare gluon propagator $\Delta_m$
with the one-loop function $-\Gamma(p)^{-1}$ inside the integrals in graphs (a1) and (a2), so that, in addition, the diagrams themselves would contain information on the position of the poles. The one-loop function $\Gamma(p^2)$ is known analytically, but the integrals would be prohibitive. On the other hand, they can be easily evaluated if the propagator is approximated by its principal part,
\BE
\Delta_P(p)= -\frac{R}{p^2-p_0^2}-\frac{R^\star}{p^2-(p_0^2)^\star}\,,
\EE
as was done in Ref.~\cite{quark} to study the infrared behavior of the quark propagator. The principal part $\Delta_P(p)$ is the largest contribution to the one-loop gluon propagator of Fig.~1, and by a slight renormalization it provides a very good approximation of the exact propagator in the IR \cite{xigauge}. It is also equivalent to the leading order propagator of the refined Gribov-Zwanziger effective theory~\cite{dudal08,capri,dudal10,dudal11,dudal18}.

In graphs (a1) and (a2), an approximate resummation of all the polarization insertions in the internal gluon line can be achieved very easily, 
without having to evaluate new integrals, by replacing $\Delta_{m}\to \Delta_{P}$ under the sign of integral and using a trick
which was discussed in Ref.~\cite{quark}. Using the linearity of the one-loop graphs, the result follows by analytic continuation of the mass
parameter $m$ in the complex plane. If we denote by $\Delta_{p_0}$ a bare propagator $\Delta_m$ with the mass $m$ replaced by $p_0$,
then the principal part can be written as
\BE
\Delta_P=\frac{1+i\tan\theta}{2}\, \Delta_{p_0}+\frac{1-i\tan\theta}{2}\,\Delta_{p_0^\star}\,,
\EE
where as before $\theta$ is the phase of the residue and the overall normalization of $\Delta_{P}$ is fixed so that $\Delta_P(p)\sim\Delta_m(p)$ in the UV, thus ensuring that the divergent part of the integral 
does not change.
Denoting by $s_0=p_0^2/m^2$ the adimensional pole position, we can then define a modified resummed logarithmic function $L^R$  as
\BE
L^R(s)=\frac{1+i\tan\theta}{2}\, L(s/s_0)+\frac{1-i\tan\theta}{2}\, L(s/s_0^\star)\,,
\label{LR}
\EE
to replace the function $L(s)$ in Eq.~(\ref{Fs2}). A rigorous proof of the procedure is given in Ref.~\cite{quark}.

\begin{figure}[b] \label{fig:contour3}
\centering
\includegraphics[width=0.35\textwidth,angle=-90]{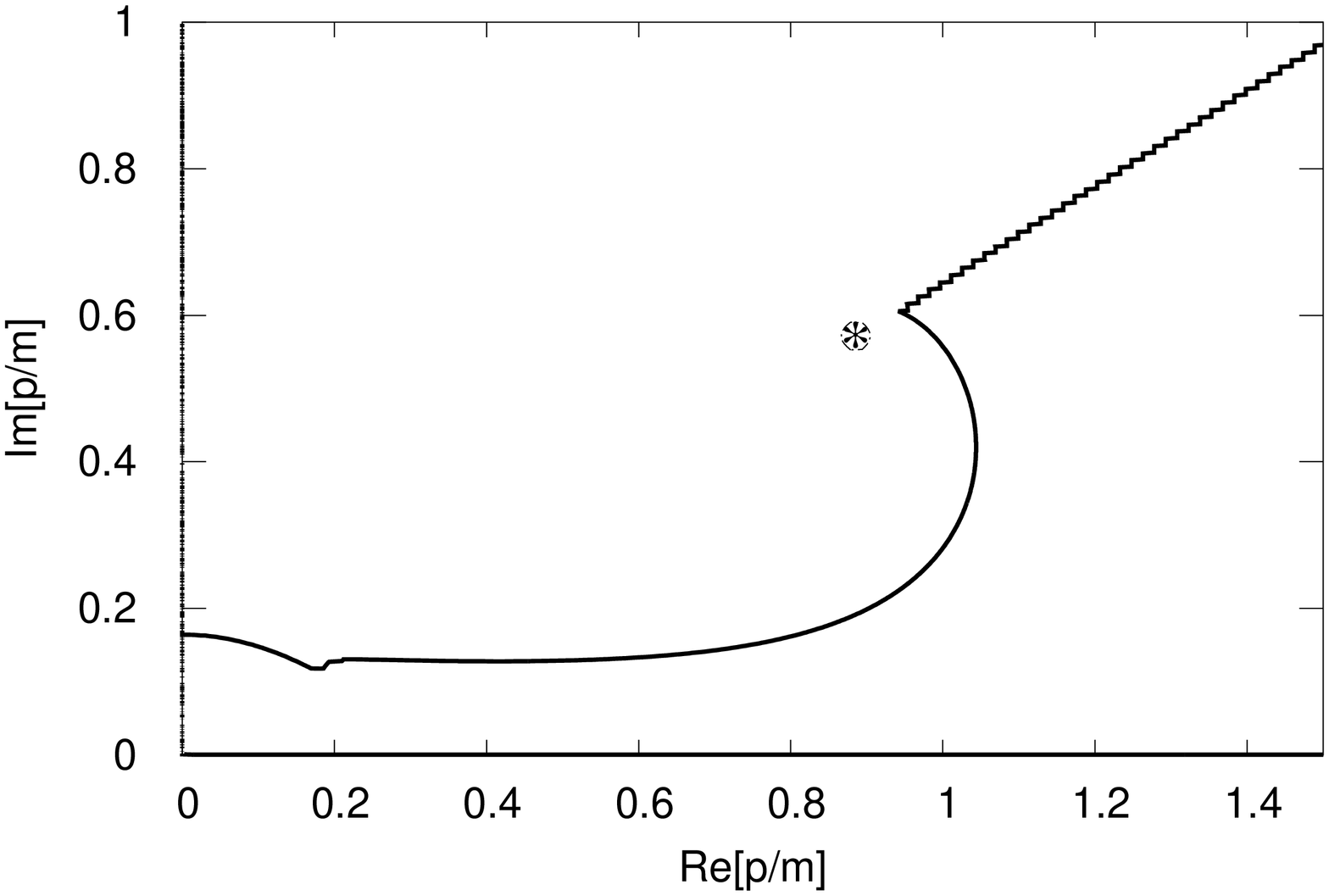}
\caption{Same as Fig.~4, but with a full resummation of the internal gluon line by the principal part, according to
Eq.~(\ref{LR}), which is used for the calculation instead of Eq.~(\ref{Ls2}). The wavy line is a branch cut where the function changes sign without crossing the zero.}
\end{figure}

\begin{figure}[b] \label{fig:log}
\centering
\includegraphics[width=0.35\textwidth,angle=-90]{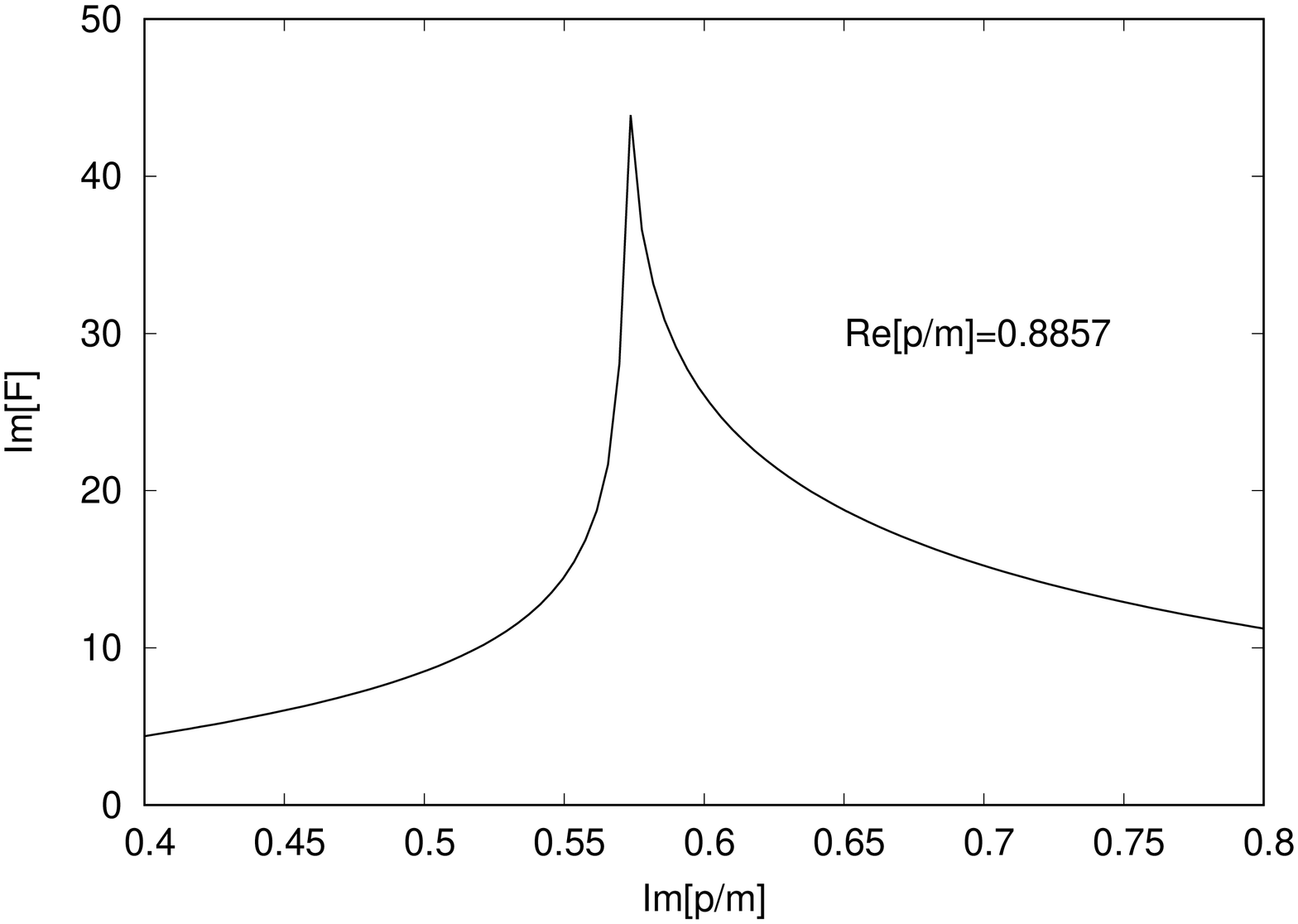}
\caption{Logarithmic divergence of the function $F$ at the pole, according to the resummation scheme of Eq.~(\ref{LR}).
The imaginary part, $\Im\{F(-p^2/m^2)\}$ is evaluated by Eq.~(\ref{LR})  as a function of $\Im(p/m)$ across the pole, with $\Rerm{p/m}$ kept fixed at the pole value, $\Rerm{p/m}=0.8857$.}
\end{figure}

The analytic properties of the function $F$ change dramatically when the internal gluon line is replaced by the principal part, with complex poles, using Eq.~(\ref{LR}) instead of Eq.~(\ref{Ls2}) in Eq.~(\ref{Fs2}).
As shown in Fig.~6, the line of zeros of the imaginary part is strongly modified and reaches a point very close to the pole. In more detail, the line of zeros merges with a branch cut which -- albeit not clearly visible in the plot -- originates at the pole itself. In the figure, the branch cut is depicted as a wavy line, along which the function changes sign
without going to zero\footnote{As such, the branch cut should not be viewed as being part of the contour $\Im\{F(-p^2/m^2)\}=0$, of course.}.

The branch cut is explained by the existence of a logarithmic divergence at the pole. In Fig.~7, this logarithmic divergence is displayed by plotting the function $\Im[F]$ with the real part $\Rerm[p/m]$ kept fixed at the pole value, $\Rerm[p_0/m]=0.8857$, while changing the imaginary part, $\Im[p/m]$, across the pole, which occurs at $\Im[p_0/m]=0.5718$.
The logarithmic divergence arises from the divergence of $L(s)$ at $s=-1$ in Eq.~(\ref{Ls2}); it follows the pole and 
moves to $s=-s_0$ in Eq.~(\ref{LR}), as $L$ gets replaced by $L^{R}$.

The divergence does not spoil the invariance of the pole, since it is anyways canceled by the zero of $\Gamma$ in Eq.~(\ref{nielsen1}). On the other hand, at the level of the derivative $\partial\theta/\partial\xi$, the phase of the residue would receive an unphysical diverging term if the logarithm were not compensated by an extra zero at the pole. In other words, if the branch cut is a true feature of the exact result, then the Nielsen identity for the phase is well-defined only if the exact line of zeros reaches the pole. We may then expect that vertex and higher order corrections, when included, would drive the imaginary part of $F$ towards an exact zero at $p_{0}$, in order to reconcile the identity with the expectation of a finite phase change for the residue. The vanishing of the derivative of the phase would then follow.

More generally, if the logarithmic divergence were genuine -- with no zero to tame it in $F$ --, the diverging phase would be the sign of a branch point at the zero of $\Gamma$. $p_{0}$ would not then be a true pole of the propagator, and the steps which led to Eqs.~(\ref{imag}) and (\ref{imag2}) would be invalidated. The Nielsen identity  would still hold, 
but its relation to the phase would be lost, because there would not be a well-defined residue in the first place. 
Of course, in this scenario, the gluon propagator would have no principal part at all, which is in contrast
with what was found by the screened expansion at one loop. This is quite unlikely, if we look at the excellent agreement
which is found with the lattice data in Fig.~1.

At the same time, it has been recently shown that without a knowledge of the exact vertex structure,
nothing can prevent a wild proliferation of unphysical branch cuts, order by order, when complex-conjugated
poles are present in the propagator~\cite{pawlowski22}. Thus, the logarithmic divergence and associated branch cut might just be a consequence of the missing resummation of vertex corrections. In the complete absence of a logarithmic divergence at $p_{0}$, $\Im[F]_{p_{0}}$ would be finite with a line of zeros passing remarkably close to it. Thus, again, our resummed results point to either the derivative $\partial\theta/\partial\xi$ being exactly zero, or at least to it being very small.

We are far from having reached a formal proof of the vanishing of the gauge-derivative of the phase, of course. Nonetheless, we argue that the exact gauge-parameter-independence of the phase $\theta$ is the only reasonable assumption which could avoid any conflict between the exact Nielsen identity and the results of the screened expansion. Were the conflict an artifact of the expansion itself, we argue that there are good indications from our results
that the derivative of the phase is at least very small, if not exactly zero.

\acknowledgments

This research was supported in part by the INFN-SIM national project and by the ``Linea di intervento 2'' for HQCDyn at DFA-Unict.

\appendix
\section{one-loop graphs}

\subsection{Graphs (a1) and (a2)}

The first graph, (a1) in Fig.~2,  reads
\begin{widetext}
\begin{align}
2{\cal F}_{ab}^{\mu\nu}(p)&=-(-ip^\mu)\left(\frac{i}{p^2}\right) p_\alpha g f_{dac}
\int\kkk
\,i\Delta_m^{\alpha\beta}(k)\left[g f_{dcb} V_{\beta\sigma\lambda}(p,k)\right]
\left[\frac{i}{(k-p)^2}\right]\left[\frac{(p-k)^\sigma}{(p-k)^2}\right]\, i\Delta_m^{\lambda\nu}(p)\,,
\end{align}
where the three-gluon vertex structure is
\BE
V_{\beta\sigma\lambda}(p,k)=g_{\beta\sigma}(p-2k)_\lambda
+g_{\sigma\lambda}(k-2p)_\beta+g_{\lambda\beta}(k+p)_\sigma\,,
\EE
while the ghost and mixed propagators are $\langle c\bar c\rangle=i/p^2$ and 
$\langle BA^\mu\rangle=p^\mu/p^2$, respectively, according to Eq.~(\ref{D0}) and Eq.~(\ref{BA}). The massive
gluon propagator $\Delta_m$ was defined in Eq.~(\ref{Deltam}). As a countercheck of sign consistence we can use the
Slavnov-Taylor identity 
$0=\langle \delta_\theta(A_\mu\bar c)\rangle=\langle (D_\mu c )\bar c\rangle+\langle A_\mu B\rangle$ leading to
$\langle A_\mu B\rangle=-\langle (\partial_\mu c )\bar c\rangle=-(-ip_\mu)(i/p^2)=-p_\mu/p^2$ as $g\to 0$.
The sum over color indices gives $f_{dac}f^{dcb}=-N\delta_{ab}$ and dropping the delta
\BE
2{\cal F}^{\mu\nu}(p)=-ig^2N\,\frac{p^\mu p_\alpha}{p^2}\
\int\kkk
\Delta_m^{\alpha\beta}(k) V_{\beta\sigma\lambda}(p,k)
\frac{(p-k)^\sigma}{(p-k)^4}\,\Delta_m^{\lambda\nu}(p)\,.
\EE
According to Eq.~(\ref{trans}), the sum of the first pair of graphs, (a1) and (a2) in Fig.~2, can be written as
\BE
2{\cal F}^{\mu\nu}_{(a1+a2)}(p)=ig^2N\, t^\mu_{\>\alpha}(p)\,\Delta_m^{\lambda\nu}(p)
\int\kkk
\Delta_m^{\alpha\beta}(k) V_{\beta\sigma\lambda}(p,k)
\frac{(p-k)^\sigma}{(p-k)^4}\,,
\EE
and by a bit of algebra
\BE
2{\cal F}^{\mu\nu}_{(a1+a2)}(p)=-g^2N\, t^\mu_{\>\alpha}(p)
\, I^\alpha_{\>\lambda}(p)
\,\Delta_m^{\lambda\nu}(p)\,,
\label{F12}
\EE
where the integral $I^\alpha_{\>\lambda}(p)$ is
\BE
 I^\alpha_{\>\lambda}(p)=
i\,\int\kkk
\frac{ \Delta_m^{\alpha\beta}(k) }{(p-k)^4}
\left[ k^2 t_{\beta\lambda}(k)-p^2t_{\beta\lambda}(p)\right]\,.
\EE
We need the transverse part of the integral to be inserted in Eq.~(\ref{F12}).  
Thus, replacing $t_{\beta\lambda}(p)$ by $g_{\beta\lambda}$ and
using the identity
\BE
\frac{1}{k^2(k^2-m^2)}=\frac{1}{m^2}\left[\frac{1}{k^2-m^2}-\frac{1}{k^2}\right]\,,
\EE
 the integral reads
\BE
 I_{\alpha\lambda}(p)=\frac{m^2-p^2}{m^2}\, J_{\alpha\lambda}(p,m)
+\frac{p^2}{m^2}\, J_{\alpha\lambda}(p,0)+\xi p^2 K_{\alpha\lambda}(p)\,,
\label{I}
\EE
where
\begin{align}
 J_{\alpha\lambda}(p,m)&=
-i\,\int\kkk
\frac{k^2 g_{\alpha\lambda}-k_\alpha k_\lambda}{\left[(k-p)^2\right]^2 (k^2-m^2)}\,,\nn\\
 K_{\alpha\lambda}(p)&=
i\,\int\kkk
\frac{k_\alpha k_\lambda}{\left[(k-p)^2k^2\right]^2}\,.
\end{align}
By Feynman parametrization and neglecting odd terms in the numerator, the first integral splits as 
\BE
 J^\alpha_{\>\lambda}(p,m)=\tilde J\,^\alpha_{\>\lambda}(p,m)+p^2t^\alpha_{\>\lambda}(p) J(p,m)\,,
\label{J}
\EE
where
\begin{align}
\tilde J\,^\alpha_{\>\lambda} (p,m)&=2\int_0^1 x {\rm d}x\int\qqqE
\frac{(q^2\delta^\alpha_\lambda -q^\alpha q_\lambda)} {(q^2+M_{x,m}^2)^3}\,,\nn\\
J(p,m)&=-2\int_0^1 x^3 {\rm d}x\int\qqqE
\frac{1} {(q^2+M_{x,m}^2)^3}\,,
\end{align}
and the integrals are in the Euclidan space where  $q_\alpha$ is defined, while the mass function $M^2_{x,m}$ is
\BE
M^2_{x,m}=(1-x)\left[m^2-xp^2\right]\,.
\EE
By the same notation, dropping a longitudinal term, the integral $K_{\alpha\lambda}$ reads
\BE
K^\alpha_{\>\lambda}(p)=(3!)\int_0^1 x(1-x) {\rm d}x\int\qqqE
\frac{q^\alpha q_\lambda} {(q^2+M_{x,0}^2)^4}\,.
\EE

The integral $\tilde J$ is evaluated by dimensional regularization with $d=4-2\epsilon$ and  an arbitrary scale factor
$\mu^{2\epsilon}$
\begin{align}
\tilde J\,^\alpha_{\>\lambda} (p,m)&=2\delta^\alpha_\lambda 
\left(\frac{d-1}{d}\right)
\int_0^1 x {\rm d}x \>
\frac{\mu^{2\epsilon}}{2^{d-1}\pi^{d/2}\Gamma (d/2)} (M_{x,m})^{d-4}\frac{\Gamma(d/2+1)\Gamma(2-d/2)}{2\Gamma(3)}\,,
\end{align}
yielding
\BE
\tilde J\,^\alpha_{\>\lambda} (p,m)=
\frac{3\delta^\alpha_\lambda }{32\pi^2}
\int_0^1 x {\rm d}x \>\left[\frac{1}{\hat \epsilon}-\frac{2}{3}-\log{\frac{M_{x,m}^2}{m^2}}\right]\,,
\EE
where the diverging part is
\BE
\frac{1}{\hat \epsilon}=\frac{1}{\epsilon}-\gamma+\log\frac{4\pi\mu^2}{m^2}\,,
\label{eps}
\EE
while $J(p,m)$ and $K^\alpha_{\>\lambda}(p)$ are finite in the UV\,,
\begin{align}
J(p,m)&=-2\int_0^1 x^3 {\rm d}x\left[
\frac{1}{32\pi^2}\frac{1} {M_{x,m}^2}\right]\,,\nn\\
K\,^\alpha_{\>\lambda} (p)&=\frac{(3!)\delta^\alpha_\lambda }{4}\int_0^1 x(1-x) {\rm d}x \>
\left[\frac{1}{(3!)\,8\pi^2}\frac{1}{ M^2_{x,0} }\right]\,.
\end{align}

The remaining integrals are elementary:
\begin{align}
\int_0^1 x {\rm d}x&=\frac{1}{2}\nn\\
\int_0^1 x \log(1-x){\rm d}x&=-\frac{3}{4}\nn\\
\int_0^1 x \log(1+sx) {\rm d}x&=\frac{1}{2}\left[ L_1(s)-\frac{1}{2}\right]\nn\\
\int_0^1 \frac{s x^3 }{(1-x)(1+sx)}  {\rm d}x&
=\frac{s}{1+s}\left[\lim_{\eta\to 0} \int_0^{1-\eta} \frac{ x^2 }{(1-x)}  {\rm d}x
-\int_0^1 \frac{ x^2 }{(1+sx)}  {\rm d}x\right]\nn\\
&=\frac{s}{1+s}\left[ L_2(s)-\frac{3}{2}-\log\eta+{\cal O}(\eta)\right]\,,
\end{align}
where the limit $\eta\to 0$ must be taken at the end of the calculation in order to deal with the spurious IR 
divergence which arises in the first integral of the last line.
The logarithmic functions $L_1$, $L_2$ are defined as
\begin{align}
L_1(s)&=\frac{1}{s}+\frac{s^2-1}{s^2}\log(1+s)\,,\nn\\
L_2(s)&=\frac{1}{s^2}-\frac{1}{2s}-\frac{1}{s^3}\log(1+s)\,,
\end{align}
where the variable $s$ is the Euclidean squared momentum $s=-p^2/m^2$ in units of the mass parameter $m$.
Inserting these explicit expressions, the integrals $\tilde J$, $J$ and $K$ read
\begin{align}
\tilde J\,^\alpha_{\>\lambda} (p,m)&=
\frac{3\delta^\alpha_\lambda }{64\pi^2}
\left[\frac{1}{\hat \epsilon}+\frac{4}{3}-L_1(s)\right]\nn\\
p^2 J(p,m)&=
\frac{2 }{32\pi^2}\left(\frac{s}{1+s}\right)\left[ L_2(s)-\frac{3}{2}-\log\eta\right]\nn\\
p^2 K^\alpha_{\>\lambda}(p)&=
-\frac{\delta^\alpha_\lambda  }{32\pi^2}\,. 
\label{JK}
\end{align} 
We observe that,  in the limit $s\to \infty$, which is equivalent to $m\to 0$, 
the logarithmic functions $L_1(s)$, $L_2(s)$ have the asymptotic behavior 
\begin{align}
L_1(s)&=\log(s)+\frac{2}{s}+{\cal O}(s^{-2})\,,\nn\\ 
L_2(s)&= -\frac{1}{2s}+{\cal O}(s^{-2})\>\to\> 0\,.
\label{asympt}
\end{align}
Then, we can write
\begin{align}
(1+s)\, \tilde J\,^\alpha_{\>\lambda} (p,m)- s\,\tilde J\,^\alpha_{\>\lambda} (p,0) &=
\delta^\alpha_\lambda\frac{3 }{64\pi^2}
\left[\frac{1}{\hat \epsilon}+\frac{4}{3}-(1+s)L_1(s)+s\log s\right]\,,\nn\\
(1+s)\,p^2 J(p,m)-s\,p^2 J(p,0) &=\frac{4 }{64\pi^2} \left[s\,L_2(s)\right]\,.
\end{align}
Because of the transverse projector in Eq.~(\ref{F12}), we can replace $t^\alpha_{\>\lambda}(p)$ by $\delta^\alpha_{\>\lambda}$
in Eq.~(\ref{J})  and insert it in Eq.~(\ref{I}) which reads
\BE
 I^\alpha_{\>\lambda}(p)=\frac{\delta^\alpha_\lambda }{64\pi^2}
\left[ \frac{3}{\hat \epsilon}-3(1+s)L_1(s)+3s\log s  +4s\, L_2(s) -2\xi+4\right]\,.
\EE
Finally, the first pair of graphs, (a1) and  (a2) in Fig.~2, give the pure transverse sum
\BE
2{\cal F}^{\mu\nu}_{(a1+a2)}(p)=\frac{g^2N}{64\pi^2} \, \frac{t^{\mu\nu}(p)}{ p^2-m^2}
\left[ \frac{3}{\hat \epsilon}-3(1+s)L_1(s)+3s\log s  +4s\, L_2(s) -2\xi+4\right].
\label{a1a2}
\EE
We notice the presence of the transverse part of the bare massive propagator $\Delta_m(p)$,
as a factor which arises from the external gluon leg. As a check, in the limit $m\to 0$, which is equivalent to $s\to \infty$,
we recover the same result -- modulo irrelevant constants -- that was found in Ref.~\cite{breck} by standard perturbation theory, namely
\BE
\left[2{\cal F}^{\mu\nu}_{(a1+a2)}(p)\right]_{m=0}=\frac{g^2N}{64\pi^2} \, \frac{t^{\mu\nu}(p)}{ p^2}
\left[ \frac{3}{\hat \epsilon}-3\log(-p^2/m^2)  -2\xi-4\right]\,.
\label{a1a2PT}
\EE

\subsection{Graphs (b1) and (b2)}

The second pair of graphs, (b1) and (b2) in Fig.~2, have no internal gluon lines and there are no masses in the internal propagators.
Thus, the result is the same as for standard perturbation theory, apart from the external bare gluon line. As a check of consistence,
here we recover the explict result of Ref.~\cite{breck} by our notation. The graph (b1) gives 
\begin{align}
2{\cal F}_{ab}^{\mu\nu}(p)&=-(-ip^\mu)\left(\frac{i}{p^2}\right) p_\alpha g f^{dac}
\int\kkk
\left[-\frac{(p-k)^\alpha}{(p-k)^2}\right]\left[\frac{i}{k^2}\right]\left[\frac{i}{(p-k)^2}\right]
\left[k_\lambda g f^{bcd}\right]
 i\Delta_m^{\lambda\nu}(p)\,,
\end{align}
and dropping the $\delta_{ab}$ which arises from the sum over color indices
\BE
2{\cal F}^{\mu\nu}(p)=-ig^2N\,\frac{p^\mu p_\alpha}{p^2}
\int\kkk
\frac{(p-k)^\alpha k_\lambda}{(p-k)^4k^2}\,\Delta_m^{\lambda\nu}(p)\,.
\EE
According to Eq.~(\ref{trans}), the sum of the second pair of graphs, (b1) and (b2) in Fig.~2, can be written as
\BE
2{\cal F}^{\mu\nu}_{(b1+b2)}(p)=ig^2N\, t^\mu_{\>\alpha}(p)\,\Delta_m^{\lambda\nu}(p)
\int\kkk
\frac{(p-k)^\alpha k_\lambda}{(p-k)^4k^2}
=g^2N\, t^\mu_{\>\alpha}(p)\,    T^\alpha_{\>\>\lambda}(p) \,  \Delta_m^{\lambda\nu}(p)\,,
\label{Fb12}
\EE
where, dropping a longitudinal term, the integral $T^\alpha_{\>\lambda}(p)$ is
\BE
 T^\alpha_{\>\>\lambda}(p)=
-i\,\int\kkk
\frac{k^\alpha k_\lambda}{(p-k)^4k^2}\,.
\EE
By Feynman parametrization and, again, neglecting odd and longitudinal terms, the integral can be evaluated in
the Euclidean space and reads
\BE
 T^\alpha_{\>\>\lambda}(p)=2\int_0^1 x {\rm d}x\int\qqqE
\frac{q^\alpha q_\lambda} {(q^2+M_{x,0}^2)^3}
= \left(\frac{1}{d-1}\right) \tilde J^\alpha_{\>\lambda} (p,0)\,.
\EE
By dimensional regularization, adding the factor $(d-1)^{-1}\approx (1/3)(1+2\epsilon/3)$ and using the
asymptotic behavior of $L_1(s)$, Eq.~(\ref{asympt}), the integral follows from the first line of Eq.~(\ref{JK}),
\BE
 T^\alpha_{\>\>\lambda}(p)=
\frac{\delta^\alpha_\lambda }{64\pi^2}
\left[\frac{1}{\hat \epsilon}+2-\log(s)\right]\,,
\label{T}
\EE
and by insertion in Eq.(\ref{Fb12}) we obtain the final result
\BE
2{\cal F}^{\mu\nu}_{(b1+b2)}(p)=-\frac{g^2N}{64\pi^2} \, \frac{t^{\mu\nu}(p)}{ p^2-m^2}
\left[ \frac{1}{\hat \epsilon}-\log(-p^2/m^2) +2\right]\,,
\label{b1b2}
\EE
which agrees with Ref.~\cite{breck}, apart from the denominator, $p^2-m^2$, which arises from the external gluon leg and is replaced by the bare denominator, $p^2$,
in the standard perturbation theory.
 
\subsection{Total one-loop contribution}

The sum of all the uncrossed one-loop graphs in Fig.~2 gives
\BE
2{\cal F}^{\mu\nu}(p)=\frac{g^2N}{64\pi^2} \, \frac{t^{\mu\nu}(p)}{ p^2-m^2}
\left[ \frac{2}{\hat \epsilon}-3L(-p^2/m^2)+\log(-p^2/m^2) -2\xi-6\right]\,,
\label{1loop}
\EE
where the logarithmic function $L(s)$ is defined as
\begin{align}
L(s)&=(1+s)L_1(s)-s\log s-2-\left[\frac{4 s}{3} L_2(s)+\frac{2}{3}\right]\nn\\
&=\frac{1+s}{s}+\frac{(1+s)(s^2-1)}{s^2}\log(1+s)-2-s\log s
-\frac{4}{3}\left[\frac{1}{s}-\frac{1}{s^2}\log(1+s)\right]
\label{Ls}
\end{align}
and has the leading behavior $L(s)\sim \log s$ in the limit $s\to \infty$ or $m\to 0$.

In the limit $m\to 0$, modulo an irrelevant constant, we recover the result of standard perturbation theory\,\cite{breck}
\BE
\left[2{\cal F}^{\mu\nu}(p)\right]_{m=0}=\frac{g^2N}{64\pi^2} \, \frac{t^{\mu\nu}(p)}{ p^2}
\left[ \frac{2}{\hat \epsilon}-2\log(-p^2/m^2)  -2\xi-6\right]\,.
\label{1loopPT}
\EE

\end{widetext}

\end{document}